\documentclass[preprint,showpacs,preprintnumbers,amsmath,amssymb]{revtex4}
\usepackage{graphicx}% Include figure files
\usepackage{epsfig}
\usepackage{dcolumn}% Align table columns on decimal point
\usepackage{bm}% bold math
\def\nA{nucleon-nucleus\ }
\def\pA{proton-nucleus\ }
\def\aA{$\alpha$-nucleus\ }
\def\AA{nucleus-nucleus\ }
\def\phe6{$^6$He+$p$\ }
\def\he6pn{$p(^6$He,$^6$Li$^*)n$\ }
\def\pn{$(p,n)$\ }
\def\pp{$(p,p)$\ }
\def\pnA{$A_{\rm g.s.}(p,n)\tilde{A}_{\rm IAS}$\ }
\def\pnCa{$^{48}$Ca$_{\rm g.s.}(p,n)^{48}$Sc$_{\rm IAS}$\ }
\def\pnZr{$^{90}$Zr$_{\rm g.s.}(p,n)^{90}$Nb$_{\rm IAS}$\ }
\def\pnSn{$^{120}$Sn$_{\rm g.s.}(p,n)^{120}$Sb$_{\rm IAS}$\ }
\def\pnPb{$^{208}$Pb$_{\rm g.s.}(p,n)^{208}$Bi$_{\rm IAS}$\ }
\def\pphe6{$p(^6$He,$^6$He)$p$\ }
\def\p2he6{$p(^6$He,$^6$He$^*)p$'\ }
\def\nli6{$^6$Li$^*+n$\ }
\def\An{$n+\tilde{A}$\ }
\def\Ap{$p+A$\ }
\def\pCa{$p+^{48}$Ca\ }
\def\pZr{$p+^{90}$Zr\ }
\begin{document}
%\preprint{Revised version submitted to Phys. Rev. C}
\title{Folding model study of the isobaric analog excitation:
isovector density dependence, Lane potential and nuclear symmetry energy}
\author{Dao T. Khoa}\email{khoa@vaec.gov.vn}
\author{Hoang Sy Than}
\author{Do Cong Cuong}
\affiliation{Institute for Nuclear Science {\rm \&} Technique, VAEC, P.O. Box
5T-160, Nghia Do, Hanoi, Vietnam.}

\date{June 8, 2007; accepted for publication in Phys. Rev. C}
%\date{\today}

\begin{abstract}
A consistent folding model analysis of the ($\Delta S=0, \Delta T=1$) charge
exchange \pn reaction measured with $^{48}$Ca, $^{90}$Zr, $^{120}$Sn and
$^{208}$Pb targets at the proton energies of 35 and 45 MeV is done within a
two-channel coupling formalism. The nuclear ground state densities given by the
Hartree-Fock-Bogoljubov formalism and the density dependent CDM3Y6 interaction
were used as inputs for the folding calculation of the nucleon optical
potential and \pn form factor. To have an accurate isospin dependence of the
interaction, a complex isovector density dependence of the CDM3Y6 interaction
has been carefully calibrated against the microscopic Brueckner-Hatree-Fock
calculation by Jeukenne, Lejeune and Mahaux before being used as folding input.
Since the isovector coupling was used to explicitly link the isovector part of
the nucleon optical potential to the cross section of \pn reaction exciting the
0$^+$ isobaric analog states in $^{48}$Sc, $^{90}$Nb, $^{120}$Sb and
$^{208}$Bi, the newly parameterized isovector density dependence could be well
tested in the folding model analysis of the \pn reaction. The isospin- and
density dependent CDM3Y6 interaction was further used in the Hartree-Fock
calculation of asymmetric nuclear matter, and a realistic estimation of the
nuclear symmetry energy has been made.
\end{abstract} \pacs{24.50.+g, 25.60.Bx, 25.60.Lg,} \maketitle

\section{Introduction}
The charge exchange \pn reaction is well known as an effective tool to excite
the isobaric analog of the target ground state. Such an isobaric analog state
(IAS) has essentially the same structure as the ground state of the target
except for the replacement of a neutron by a proton and, hence, differs in
energy approximately by the Coulomb energy of the added proton. In the isospin
representation, the two states are just two members of the isospin multiplet
which differ only in the orientation of the isospin $\bm{T}$. The similarity of
the initial and final states of the \pn reaction makes this reaction very much
like an elastic scattering  in which the isospin of the incident proton is
``flipped". Indeed, the isospin dependent part of the \pA optical potential
(OP) was used by Satchler {\sl et al.} \cite{Sat64} some 40 years ago as the
charge exchange form factor in their study of the \pn reaction within the
distorted wave Born approximation (DWBA). In general, the central \nA OP can be
written in terms of the isovector coupling \cite{La62} as
\begin{equation}
 U(R)=U_0(R)+4U_1(R)\frac{{\bm t}.{\bm T}}{A}, \label{e1}
\end{equation}
where ${\bm t}$ is the isospin of the incident nucleon and ${\bm T}$ is that of
the target $A$. The second term of Eq.~(\ref{e1}), known as the Lane potential,
contributes to the elastic ($p,p$) and ($n,n$) scattering as well as to the
charge exchange ($p,n$) reaction \cite{Sat83}. While the relative contribution
by the Lane potential $U_1$ to the elastic nucleon scattering cross section is
small and amounts only a few percent for a neutron-rich target
\cite{Kho02,Kho03}, it determines entirely the (Fermi-type) $\Delta J^\pi=0^+$
transition strength of the \pn reaction to isobaric analog states. Therefore,
the \pn reaction can be used as a reliable probe of the isospin dependence of
the \pA OP.

The \nA OP has been studied over the years and there are several ``global" sets
of the OP parameters deduced from the extensive optical model analyses of
nucleon elastic scattering, like that by Becchetti and Greenlees \cite{BG69},
the CH89 global OP \cite{Va91}, and a recent systematics by Koning and
Delaroche \cite{Kon03} which covers a wide range of energies (from 1 keV to 200
MeV) and target masses ($24\leq A\leq 209$). Although parameterized in the
empirical Woods-Saxon form, these global systematics are very valuable in
predicting the \nA OP when elastic scattering data are not available or cannot
be measured which is the case for the unstable, dripline nuclei. Given a large
neutron excess in the unstable neutron-rich nuclei, it is important to know as
accurate as possible the isospin dependence of the \nA OP before using it in
various studies of nuclear reactions and nuclear astrophysics. We recall here
the two main methods used so far to determine the isospin dependence of the \nA
OP:

(i) study the elastic scattering of proton and neutron from the same target and
measured at about the same energy, where the isovector term of Eq.~(\ref{e1})
has the same strength, but opposite signs for $(p,p)$ and $(n,n)$ elastic
scattering;

(ii) study the (isospin-flip) charge exchange \pn transition between isobaric
analog states.

Since there are not sufficient high-quality \pn data available for a wide range
of target masses and proton energies, the empirical isospin dependence of the
\nA OP has been deduced \cite{BG69,Va91,Kon03} based mainly on method (i).
While these three global nucleon optical potentials have been widely used in
predicting the \nA OP in numerous direct reaction analyses within the DWBA or
coupled-channel (CC) formalism, their isospin dependence has been rarely used
to study the charge exchange \pn transition between the IAS's. The
(phenomenological) Lane potential $U_1$ has been studied in details so far at
some particular energies only, like the systematics for $U_1$ deduced from IAS
data of the \pn reaction measured at 22.8 \cite{Car75} and 35 MeV \cite{Jon00}.
Therefore, it is necessary to have a reliable microscopic prediction for $U_1$
by the folding model, to reduce the uncertainty associated with the isospin
dependence of the \nA OP.

Another very interesting  microscopic aspect of the Lane potential is that it
provides a direct link between the isospin dependence of the in-medium
nucleon-nucleon ($NN$) interaction and the charge exchange  \pn reaction, so
that accurately measured \pn cross section can serve as a good probe of the
isospin dependence of the $NN$ interaction \cite{Doe75} if the wave functions of
the involved nuclear states are known. Furthermore, within the frame of
many-body calculation of nuclear matter (NM), the asymmetry of the NM equation
of state (EOS) depends entirely on the density- and isospin dependence of the
$NN$ interaction \cite{Kho96,Zuo99}. This asymmetry is actually determined by
the NM symmetry energy $S(\rho)$ which is defined in terms of the NM binding
energy $B(\rho,\delta)$ as
\begin{equation}
B(\rho,\delta)=B(\rho,0)+S(\rho)\delta^2+O(\delta^4)+...
 \label{e2}
\end{equation}
where $\delta=(\rho_n-\rho_p)/\rho$ is the neutron-proton asymmetry parameter.
The contribution of $O(\delta^4)$ and higher-order terms in Eq.~(\ref{e2}),
i.e., the deviation from the parabolic law was proven to be negligible
\cite{Kho96,Zuo99}. The knowledge about the nuclear EOS (\ref{e2}) is well
known to be vital for the understanding of the dynamics of supernova explosion
and neutron star formation \cite{Bet90,Swe94,Ste05}.  The NM symmetry energy
determined at the saturation density, $E_{\rm sym}=S(\rho_0)$ with
$\rho_0\approx 0.17$ fm$^{-3}$, is widely known in the literature as the
\emph{symmetry energy} or symmetry coefficient. Although numerous nuclear
many-body calculations have predicted $E_{\rm sym}$ to be around 30 MeV
\cite{Kho96,Zuo99,Bra85,Pea00}, a direct experimental determination of $E_{\rm
sym}$ still remains a challenging task. One needs, therefore, to relate $E_{\rm
sym}$ to some experimentally inferrable quantity like the neutron skin in
neutron-rich nuclei \cite{Bro00,Hor01,Fur02,Die03} or fragmentation data of the
heavy-ion (HI) collisions involving $N\neq Z$ nuclei \cite{Ono03,She04,Che05}.
In our recent study of the IAS excitation in the \he6pn reaction using the
folded Lane potential $U_1$ for the charge exchange form factor \cite{Kho05},
we have shown how the NM symmetry energy can be linked to the charge exchange
\pn transition strength and, hence, be probed in the folding model analysis of
the \pn reaction. To extend the folding model study of the \pn reaction to
heavier targets to validate the conclusion made in Ref.~\cite{Kho05} for the NM
symmetry energy, we have studied in the present work the quasi-elastic \pn
scattering measured by the MSU group for $^{48}$Ca, $^{90}$Zr, $^{120}$Sn, and
$^{208}$Pb targets at the incident proton energies of 35 and 45 MeV
\cite{Doe75}. For a detailed probe of the isospin dependence of the in-medium
$NN$ interaction, a (complex) isospin- and density dependence of the CDM3Y6
interaction \cite{Kho97} has been carefully parameterized based on the
Brueckner-Hatree-Fock (BHF) calculation of nuclear matter by Jeukenne, Lejeune
and Mahaux \cite{Je77}. While the isovector part of the \nA OP in the NM limit
has been investigated in numerous BHF studies (see, e.g., Ref.~\cite{Zuo06} and
references therein), the isospin dependence predicted by such BHF calculations
was rarely tested in the DWBA or CC analysis of the charge exchange reaction to
isobaric analog states. Our present folding model study provides, therefore, an
important method to link the BHF results to the descriptions of the
quasi-elastic \pn reaction. By using the Lane potential $U_1$ to construct the
charge exchange \pn form factor based on the isospin coupling, it is also
straightforward to probe the isospin dependence of existing global \nA OP
\cite{BG69,Va91,Kon03}. In the present work, the description of the considered
\pn reactions by the three global nucleon optical potentials
\cite{BG69,Va91,Kon03} has been given, with a detailed comparison between the
results given by the CH89 potential \cite{Va91} and those of the folding model
analysis.

\section{IAS excitation, Lane potential and isospin coupling}
\label{sec2}
\subsection{General formalism}
We give here a brief introduction to the coupled-channel formalism for the
charge exchange \pn reaction to isobar analog states, and interested readers
are referred to Ref.~\cite{Sat83} for more technical details. Let us restrict
our consideration to a given isospin multiplet with fixed values of isospin
$\bm{t}$ for the projectile and $\bm{T}$ for the target. Then, the isospin
projections are $T_z=(N-Z)/2$ and $\tilde{T_z}=T_z-1$ for the target nucleus
$A$ and \emph{isobaric analog nucleus} $\tilde{A}$, respectively. We further
denote, in the isospin representation, state formed by adding a proton to $A$
as $|pA>$ and adding a neutron to $\tilde{A}$ as $|n\tilde{A}>$. The transition
matrix elements of the isovector part of the \nA optical potential (\ref{e1})
can then be obtained \cite{Sat83} for the elastic \nA scattering as
\begin{equation}
 <\tau A|4U_1(R)\frac{{\bm t}.{\bm T}}{A}|\tau A>=\pm\frac{2}{A}T_zU_1(R),
 \ {\rm with}\ \tau=p,n.
 \label{e3a}
\end{equation}
The + sign in the right-hand side of Eq.~(\ref{e3a}) pertains to incident
neutron and - sign to incident proton. Similarly, the transition matrix element
or \pn form factor (FF) for the ($\Delta T=1$) charge exchange \pnA reaction is
obtained as
\begin{equation}
 <n\tilde{A}|4U_1(R)\frac{{\bm t}.{\bm T}}{A}|pA>\equiv
 F_{pn}(R)=\frac{2}{A}\sqrt{2T_z}U_1(R).
 \label{e3b}
\end{equation}
In the two-channel approximation for the charge exchange \pn reaction to IAS,
the total wave function is written as
\begin{equation}
\Psi=|pA>\chi_{pA}({\bm R})+|n\tilde{A}>\chi_{n\tilde{A}}({\bm R}),
 \label{e4}
\end{equation}
where the waves $\chi({\bm R})$ describe the relative motion of the scattering
system. Then, the elastic $(p,p)$ scattering and charge exchange \pnA cross
sections can be obtained from the solutions of the following coupled-channel
equations \cite{Sat83}
\begin{eqnarray}
\left[K_p+U_p(R)-E_p\right]
 \chi_{pA}({\bm R})=-F_{pn}(R)\chi_{n\tilde{A}}({\bm R}),
  \label{e5a} \\
\left[K_n+U_n(R)-E_n\right]
 \chi_{n\tilde{A}}({\bm R})=-F_{pn}(R)\chi_{pA}({\bm R}).
 \label{e5b}
\end{eqnarray}
Here $K_{p(n)}$ and $E_{p(n)}$ are the kinetic-energy operators and
center-of-mass energies of the \Ap and \An partitions. The OP in the entrance
(\Ap) and outgoing (\An) channels are determined explicitly through the
isoscalar ($U_0$) and isovector ($U_1$) parts of the nucleon optical potential
(\ref{e1}) as
\begin{eqnarray}
U_p(R)& = & U_0(R)-\frac{2}{A}T_zU_1(R), \label{e6a} \\
U_n(R)& = & U_0(R)+\frac{2}{A}(T_z-1)U_1(R). \label{e6b}
\end{eqnarray}
In the CC calculation, both $U_p$ and $U_n$ are added by the corresponding
spin-orbital potential as well as $U_p$ added by the Coulomb potential of the
\Ap system. Since the energies of isobar analog states are separated
approximately by the Coulomb displacement energy, the \pn transition between
them has a nonzero $Q$ value. To account for this effect, the isoscalar $U_0$
and isovector $U_1$ potentials used to construct $F_{pn}(R)$ and $U_n(R)$ are
evaluated at an effective incident energy of $E=E_{\rm lab}-Q/2$, midway
between the energies of the incident proton and emergent neutron, as suggested
by Satchler {\sl et al.} \cite{Sat64}.

Since the existing global OP parameters \cite{BG69,Va91,Kon03} can be used to
construct the isoscalar and isovector components of the \pA OP at the
considered energies, it is straightforward to test those parameters in the
description of the \pnA reaction to isobaric analog states. However, more
interesting structure information can be obtained when $U_{0(1)}$ are evaluated
microscopically using an appropriate folding approach \cite{Kho02}.

\subsection{Folding model}
In our version \cite{Kho02,Kho03} of the single-folding model, the central \nA
potential $V$ is evaluated as a Hartree-Fock-type potential
\begin{equation}
  V=\sum_{j\in A}[<ij|v_{\rm D}|ij>+<ij|v_{\rm EX}|ji>], \label{e7}
\end{equation}
where $v_{\rm D}$ and $v_{\rm EX}$ are the direct and exchange parts of the
effective $NN$ interaction between the incident nucleon $i$ and nucleon $j$
bound in the target $A$. The antisymmetrization of the \nA system is done by
taking into account the knock-on exchange effects. To separate the isovector
part of $V$ which gives rise to the Lane potential, one needs to make explicit
the isospin degrees of freedom. Namely, the following spin-isospin
decomposition of the (energy- and density dependent) $NN$ interaction is used
\begin{eqnarray}
v_{\rm D(EX)}(E,\rho,s)=v^{\rm D(EX)}_{00}(E,\rho,s)+
 v^{\rm D(EX)}_{10}(E,\rho,s)(\bm{\sigma\sigma}') \nonumber\\
  +  v^{\rm D(EX)}_{01}(E,\rho,s)(\bm{\tau\tau}')+
 v^{\rm D(EX)}_{11}(E,\rho,s)(\bm{\sigma\sigma}')(\bm{\tau\tau}'),
\label{e8}
\end{eqnarray}
where $s$ is the internucleon distance and $\rho$ is the nuclear density around
the interacting nucleon pair. The contribution from the spin dependent terms
($v_{10}$ and $v_{11}$) in Eq.~(\ref{e8}) to the central \nA potential
(\ref{e7}) is exactly zero for a spin-saturated target like those considered in
the present work.

Using the explicit proton ($\rho_p$) and neutron ($\rho_n$) densities in the
folding input, the \nA potential (\ref{e7}) can be obtained \cite{Kho02}  in
terms of the isoscalar ($V_{\rm IS}$) and isovector ($V_{\rm IV}$) parts as
\begin{equation}
 V(E,\bm{R})=V_{\rm IS}(E,\bm{R})\pm V_{\rm IV}(E,\bm{R}),
\label{e9}
\end{equation}
where the + sign pertains to incident neutron and - sign to incident proton.
Each term in Eq.~(\ref{e9}) consists of the corresponding direct and exchange
potentials
\begin{eqnarray}
 V_{\rm IS}(E,\bm{R})=\int\{[\rho_n(\bm{r})+\rho_p(\bm{r})]
 v^{\rm D}_{00}(E,\rho,s) \nonumber\\
  +  [\rho_n(\bm{R},\bm{r})+\rho_p(\bm{R},\bm{r})]
 v^{\rm EX}_{00}(E,\rho,s)j_0(k(E,R)s)\}d^3r,
\label{e10}
\end{eqnarray}
\begin{eqnarray}
 V_{\rm IV}(E,\bm{R})=\int\{[\rho_n(\bm{r})-\rho_p(\bm{r})]
 v^{\rm D}_{01}(E,\rho,s) \nonumber\\
  +  [\rho_n(\bm{R},\bm{r})-\rho_p(\bm{R},\bm{r})]
 v^{\rm EX}_{01}(E,\rho,s)j_0(k(E,R)s)\}d^3r, \label{e11}
\end{eqnarray}
where $\rho(\bm{r},\bm{r}')$ is one-body density matrix of the target nucleus
with $\rho(\bm{r})\equiv\rho(\bm{r},\bm{r}), j_0(x)$ is the zero-order
spherical Bessel function, and the local momentum of relative motion $k(E,R)$
is determined from
\begin{equation}
 k^2(E,R)=\frac{2\mu}{{\hbar}^2}[E_{\rm c.m.}-V(R)-V_{\rm C}(R)].
\label{e5}
\end{equation}
Here, $\mu$ is the nucleon reduced mass, $V(R)$ and $V_{\rm C}(R)$ are,
respectively, the central nuclear and Coulomb potentials in the entrance
channel ($V_{\rm C}\equiv 0$ for the neutron-nucleus system). More details of
the folding calculation of $V_{\rm IS}$ and $V_{\rm IV}$ can be found in
Ref.~\cite{Kho02}.

We have further used in the folding calculation the density dependent CDM3Y6
interaction \cite{Kho97} which is based on the original M3Y interaction deduced
from the G-matrix elements of the Paris $NN$ potential \cite{Ana83}. The
density dependence of the \emph{isoscalar} part of the CDM3Y6 interaction was
introduced earlier in Ref.~\cite{Kho97} and its parameters have been carefully
tested in numerous folding model analyses \cite{Kho97,Kho95} of the elastic,
refractive \AA and \aA scattering. Since the \emph{isovector} part of the
interaction can be probed in a folding model analysis of the charge exchange
reaction only, we have developed in the present work an accurate procedure to
parameterize the isovector density dependence of the CDM3Y6 interaction based
on the BHF results for the energy and density dependent nucleon OP in nuclear
matter by Jeukenne, Lejeune and Mahaux (JLM) \cite{Je77}. The details of the
new treatment of the isovector density dependence of the CDM3Y6 interaction are
discussed in Sec.~\ref{sec4} below.

Given the isovector folded potential (\ref{e11}) determined entirely by the
neutron-proton difference in the nuclear densities, it is necessary to have the
nuclear densities determined as accurate as possible for a good prediction of
the Lane potential. In the present work we have used for the considered targets
the microscopic ground-state densities given by the Hartree-Fock-Bogoljubov
approach \cite{Gr01} where the single particle basis includes also the
continuum states. All the results of the optical model (OM) analysis of elastic
\nA scattering and CC calculation of the \pnA reaction have been obtained with
the CC code ECIS97 written by Raynal \cite{Ra97}.

\section{Prediction by the global optical potential}
 \label{sec3}
\begin{figure*}[htb]
 \vspace*{-1cm}
 \mbox{\epsfig{file=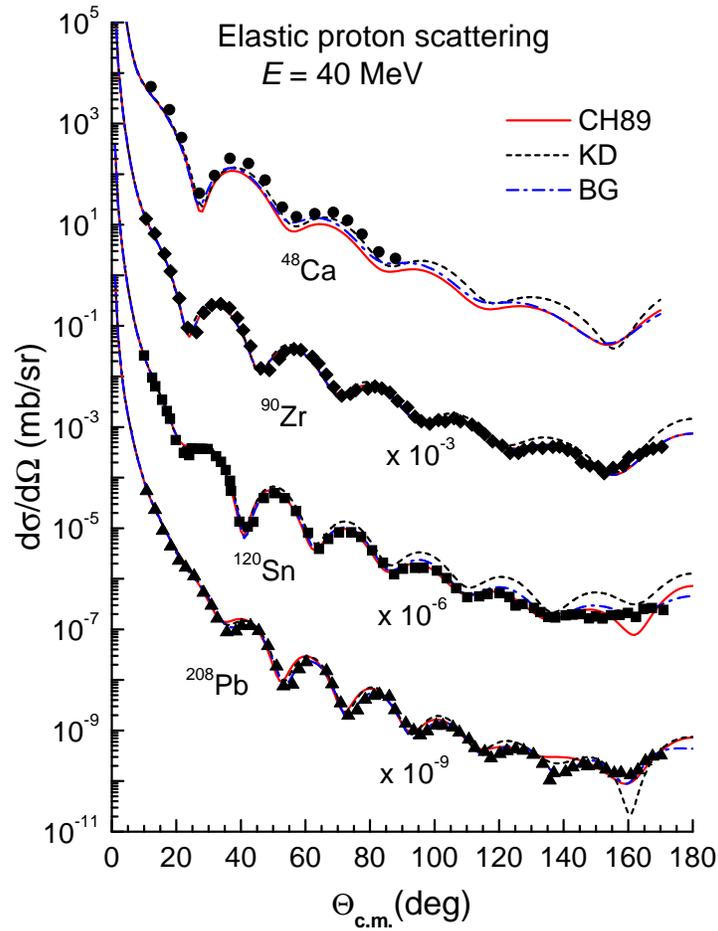,height=15cm}}\vspace*{-1cm}
\caption{(Color online) OM description of the elastic proton scattering from
$^{48}$Ca, $^{90}$Zr, $^{120}$Sn, and $^{208}$Pb targets at 40 MeV obtained
with the global OP by Becheetti and Greenlees (BG) \cite{BG69}, by Varner {\sl
et al.} (CH89) \cite{Va91}, and by Koning and Delaroche (KD) \cite{Kon03}. The
data were taken from Refs.~\cite{Gru72,Fri67}.} \label{f1}
\end{figure*}

To study the \pn reaction based on CC equations (\ref{e5a})-(\ref{e5b}), one
needs to determine the nucleon OP in the entrance ($U_p$) and outgoing ($U_n$)
channels as accurate as possible. Since the elastic neutron scattering on a
target being in its \emph{excited} IAS cannot be measured (most of IAS's are
either a short-lived bound state or an unbound resonance), we have determined
$U_n$ from the isoscalar $U_0$ and isovector $U_1$ parts of the \pA OP
evaluated at the effective incident energy $E=E_{\rm lab}-Q/2$, using
Eq.~(\ref{e6b}). The existing \nA global OP's \cite{Va91,Kon03,BG69} have been
carefully determined based on large experimental databases of both the elastic
\nA scattering and analyzing power angular distributions, and it is natural to
use them to construct $U_p$ for our study. The OM description of the elastic
proton scattering from $^{48}$Ca, $^{90}$Zr, $^{120}$Sn, and $^{208}$Pb targets
at incident proton energy of 40 MeV given by the three global \pA OP's are
shown in Fig.~\ref{f1} together with the measured data \cite{Gru72,Fri67}.
Except some underestimation of the calculated elastic cross section in \pCa
case, the overall OM description of the considered elastic scattering data is
reasonable. It should be noted that the isovector strength of the \nA OP is
only about 2-3\% of the total OP and its contribution to the elastic scattering
cross section is too weak to allow us to probe the isospin dependence of the OP
directly in the OM analysis of elastic scattering.
\begin{figure*}[htb] \vspace*{-1cm}
 \mbox{\epsfig{file=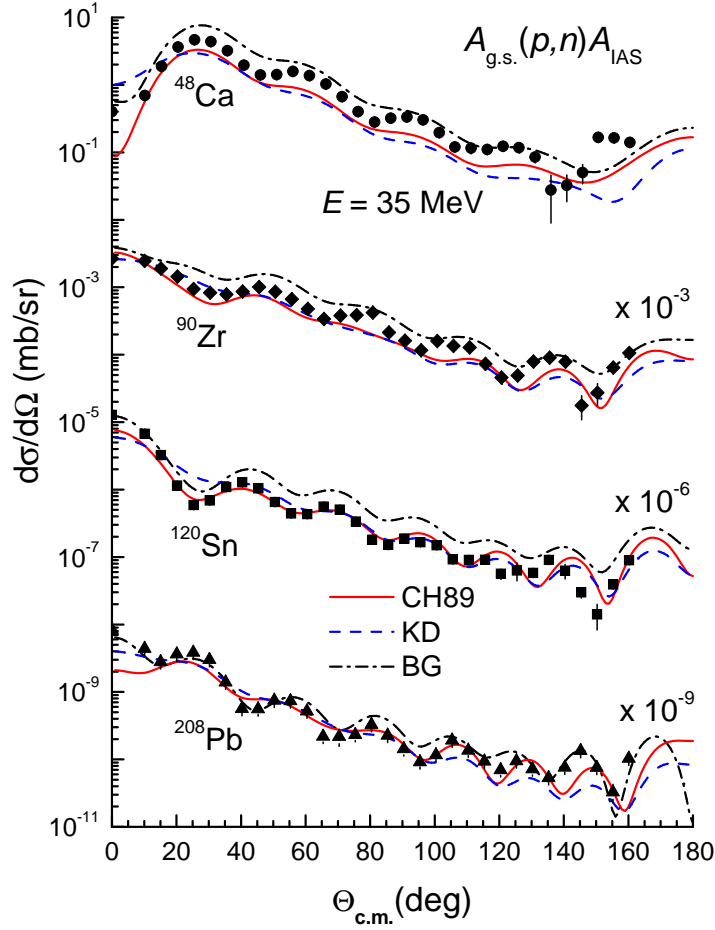,height=15cm}}\vspace*{-1cm}
\caption{(Color online) CC description of the charge exchange \pnA reaction
measured with $^{48}$Ca, $^{90}$Zr, $^{120}$Sn, and $^{208}$Pb targets at 35
MeV obtained with $U_1$ deduced from the global OP by Becheetti and Greenlees
(BG) \cite{BG69}, by Varner {\sl et al.} (CH89) \cite{Va91}, and by Koning and
Delaroche (KD) \cite{Kon03}. The data were taken from Ref.~\cite{Doe75}.}
\label{f2}
\end{figure*}
Therefore, in a ``Lane consistent" approach, the only probe of isospin
dependence of the \nA OP is the charge exchange \pnA reaction to IAS. In such a
quasi-elastic scattering, the charge exchange form factor (\ref{e3b}) used in
the CC equations (\ref{e5a})-(\ref{e5b}) is determined entirely by the Lane
potential $U_1$. As a result, any variation of the $U_1$ strength and shape can
sizably affect the calculated \pn cross section. Although all the three global
OP's give about the same OM description of the elastic proton scattering as
shown in Fig.~\ref{f1}, their descriptions of the charge exchange \pnA reaction
are quite different (see Figs.~\ref{f2} and \ref{f3}).
\begin{figure*}[htb] \vspace*{-1cm}
 \mbox{\epsfig{file=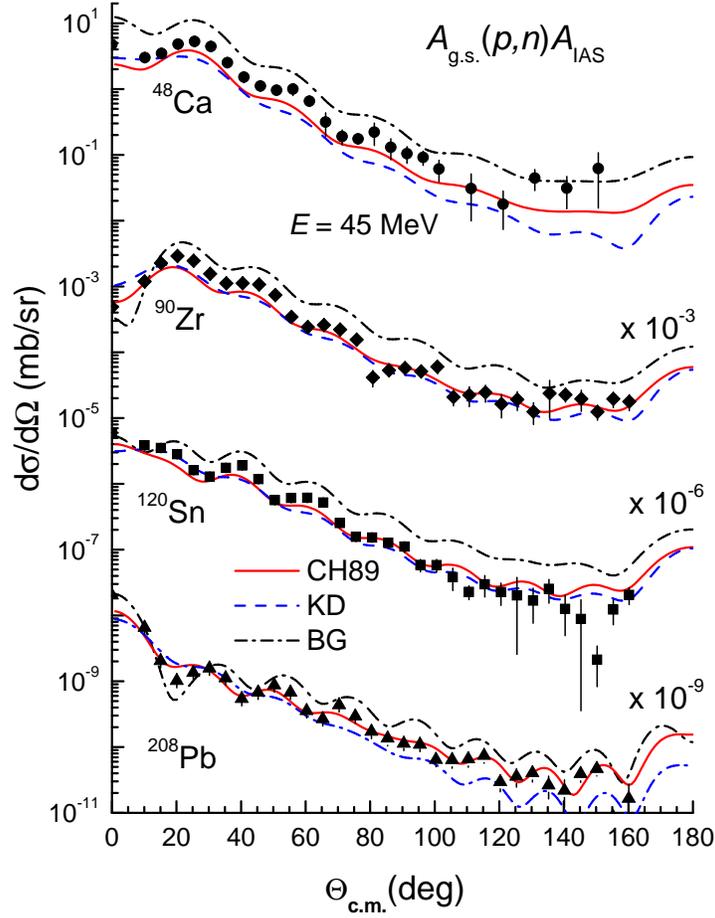,height=15cm}}\vspace*{-1cm}
 \caption{(Color online) the same as Fig.~\ref{f2} but for the \pn data measured
 at the proton energy of 45 MeV.} \label{f3}
\end{figure*}

As discussed above, the isospin dependence of the nucleon global OP
\cite{BG69,Va91,Kon03} has been determined from systematic OM studies of the
elastic scattering of proton and neutron from the same target (at about the
same energy), without any link to the \pnA reaction. Given the absolute \pn
cross section at least one order of magnitude weaker than the elastic \pp cross
section, some discrepancy in the CC description of the \pn reaction using the
Lane FF seen in Figs.~\ref{f2}-\ref{f3} is not unexpected. From the three
global OP's, $U_1$ determined from the systematics by Becheetti and Greenlees
(BG) \cite{BG69} is energy independent, and we found it too strong for the
strength of the charge exchange form factor (\ref{e3b}), especially, at energy
of 45 MeV. Such a deficiency of the BG parameters for $U_1$ was also found in
the extensive OM analysis of elastic \nA scattering \cite{Va91,Kon03}. The
isovector parts of both the global optical potentials by Varner {\sl et al.}
(CH89) \cite{Va91}, and by Koning and Delaroche (KD) \cite{Kon03} were found to
be energy dependent and weaker than that given by the BG systematics. Although
the KD global OP is more recent and covers a much wider range of energies and
target masses, from the CC results shown in Figs.~\ref{f2}-\ref{f3} one can see
that the description of \pnA reaction by the KD global OP is slightly worse
than that given by the CH89 global OP. A plausible reason is that the CH89
systematics was developed \cite{Va91} with less constraints, based only on the
elastic scattering data for $A\approx 40 - 209$ and energies of 16 to 65 MeV
(for proton) and 10 to 26 MeV (for neutron). Although this range of energies
and target masses is narrower than that covered by the KD global systematics
\cite{Kon03}, it includes the \pA systems considered in the present work. In
general, the Lane form factor (\ref{e3b}) determined from the CH89 global OP
gives a reasonable description of the \pnA cross sections measured for
$^{120}$Sn and $^{208}$Pb targets, and slightly underestimates the data for
$^{48}$Ca and $^{90}$Zr targets. As will be shown below, such a suppression of
the calculated \pnA cross sections for the two lighter targets is due mainly to
an enhanced absorption given by the CH89 global OP.

\section{Folding model analysis}
 \label{sec4}
\subsection{Isovector density dependence of the CDM3Y6 interaction}
While the isoscalar density dependence of the CDM3Y6 interaction has been well
tested in the folding model analysis \cite{Kho97,Kho95} of the elastic,
refractive \aA and \AA scattering, its \emph{isovector} density dependence can
be probed in the study of the charge exchange reaction only. In a recent work
in this direction \cite{Kho05}, we have used the same functional form for both
the isoscalar and isovector density dependences and then fine tuned the scaling
factor of the isovector part to fit the calculated \pn cross section to the
data. Although we could reach good description of the \pn reaction under study
\cite{Kho05}, it remains desirable to have a more accurate assumption for the
isovector density dependence based on the microscopic many-body calculation of
nuclear matter. Therefore, we have developed in the present work a compact
method to construct the isovector density dependence of the CDM3Y6 interaction
based essentially on the BHF description of the nucleon OP in nuclear matter by
Jeukenne, Lejeune and Mahaux \cite{Je77}. We recall that the isoscalar density
dependence of the CDM3Y6 interaction was introduced \cite{Kho97} as
\begin{eqnarray}
 v^{\rm D(EX)}_{00}(E,\rho,s)=g(E)F_{\rm IS}(\rho)v^{\rm D(EX)}_{00}(s),
\label{g1} \\
 F_{\rm IS}(\rho)=C_0[1+\alpha_0\exp(-\beta_0\rho)-\gamma_0\rho].
\label{g2}
\end{eqnarray}
Parameters of the isoscalar density dependence $F_{\rm IS}(\rho)$ were chosen
\cite{Kho97} to reproduce the NM saturation properties, with a nuclear
incompressibility $K\approx 252$ MeV, in the Hartree-Fock (HF) calculation of
symmetric NM. These parameters as well as those corresponding to other $K$
values can be found in a recent review on the \AA potential \cite{Kho07}. The
`intrinsic' energy dependence of the isoscalar interaction is contained in the
linear factor $g(E)\approx 1-0.0026E$, where $E$ is the energy of incident
nucleon. Given the success of the parametrization (\ref{g1})-(\ref{g2}) in
numerous folding calculations, we have assumed in the present work a similar
form for the isovector density dependence of the CDM3Y6 interaction
\begin{equation}
 v^{\rm D(EX)}_{01}(E,\rho,s)=F_{\rm IV}(E,\rho)v^{\rm D(EX)}_{01}(s).
 \label{g4}
\end{equation}
The radial shapes of the isoscalar and isovector interactions were kept
unchanged, as derived \cite{Kho96} from the M3Y-Paris interaction \cite{Ana83},
in terms of three Yukawas
\begin{equation}
 v^{\rm D(EX)}_{00(01)}(s)=\sum_{\nu=1}^3 Y^{\rm D(EX)}_{00(01)}(\nu)
 \frac{\exp(-R_\nu s)}{R_\nu s}.
\label{g5}
\end{equation}
One can see from the Yukawa strengths tabulated in Table~\ref{t1} that the
exchange terms $Y^{\rm EX}_{01}$ of the isovector interaction are much stronger
than the direct terms $Y^{\rm D}_{01}$ (which is due to a cancellation between
the even- and odd-state components). Therefore, an accurate evaluation of the
exchange part of the isovector potential (\ref{e11}) is very essential in the
folding model analysis of the \pn reaction. Such a domination of the exchange
term in the isovector interaction has been first noted by Love \cite{Love77}.
In our folding approach \cite{Kho02} the exchange parts of both the isoscalar
(\ref{e10}) and isovector (\ref{e11}) \pA potentials are evaluated using the
\emph{finite-range} exchange interaction $v^{\rm EX}_{00(01)}(s)$ and are,
therefore, more accurate than those given by a zero-range approximation for the
exchange term.
\begin{table}\small
\caption{\small Yukawa strengths of the central components of the M3Y-Paris
interaction (\ref{g5}).} \label{t1}
\begin{tabular}{|c|c|c|c|c|c|} \hline
$\nu$ & $R_\nu$ & $Y^{\rm D}_{00}(\nu)$ & $Y^{\rm D}_{01}(\nu)$ & $Y^{\rm
EX}_{00}(\nu)$ &
$Y^{\rm EX}_{01}(\nu)$  \\
  & (fm$^{-1}$) & (MeV) & (MeV) & (MeV) & (MeV)  \\ \hline
 1 & 4.0 & 11061.625 & 313.625 & -1524.25 & -4118.0  \\
 2 & 2.5 & -2537.5 & 223.5 & -518.75 & 1054.75  \\
 3 & 0.7072 & 0.0 & 0.0 & -7.8474 & 2.6157  \\ \hline
\end{tabular}
\end{table}

Since the nucleon OP in nuclear matter can be defined \cite{Brieva77,Huef72} as
the \emph{antisymmetrized} matrix elements of the effective $NN$ interaction
between the incident nucleon and those bound in the Fermi sea, it is given by
the same Hartree-Fock-type potential (\ref{e7}), but using \emph{plane waves}
for the single-nucleon states \cite{Kho93,Walecka}. To determine the isovector
density dependence, we have further adjusted the nucleon OP obtained with the
CDM3Y6 interaction (in the NM limit) to reproduce the JLM density- and isospin
dependent nucleon OP \cite{Je77}. Since the JLM potential is \emph{complex}, we
have used two different CDM3Y functionals to match separately the \emph{real}
and \emph{imaginary} parts of the isovector CDM3Y6 potential to those of the
JLM potential. Namely,
\begin{equation}
 F^u_{\rm IV}(E,\rho)=C^u_1(E)[1+\alpha^u_1(E)
 \exp(-\beta^u_1(E)\rho)-\gamma^u_1(E)\rho],
 \label{c1}
\end{equation}
so that the real ($u=V$) and imaginary ($u=W$) parts of the isovector CDM3Y6
interaction are determined as
\begin{eqnarray}
 v^{\rm D(EX)}_{01}(E,\rho,s)=F^{\rm V}_{\rm IV}(E,\rho)v^{\rm D(EX)}_{01}(s),
 \label{c2} \\
 w^{\rm D(EX)}_{01}(E,\rho,s)=F^{\rm W}_{\rm IV}(E,\rho)v^{\rm D(EX)}_{01}(s).
 \label{c3}
\end{eqnarray}
\begin{figure*}[htb] \vspace*{-0.5cm}
\mbox{\epsfig{file=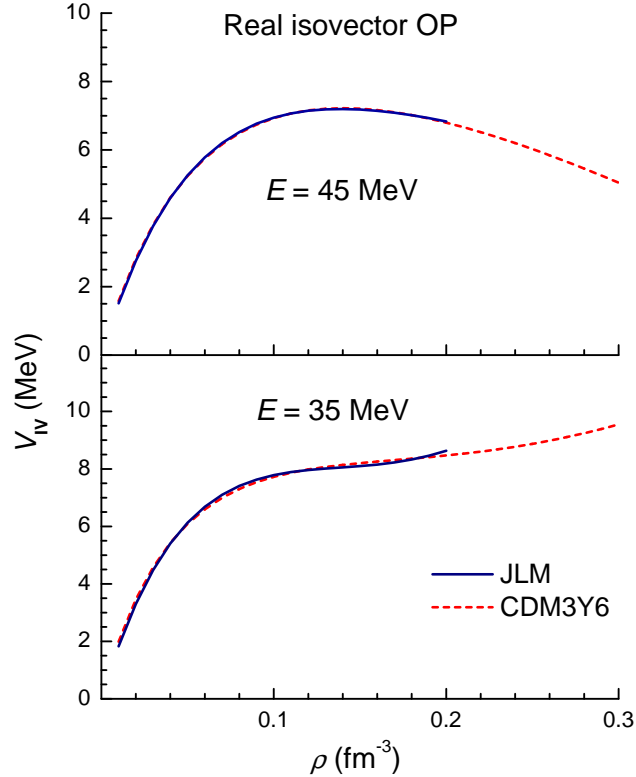,height=13cm}}\vspace*{-1.5cm} \caption{(Color
online) real part $V_{\rm IV}(E,\rho)$ of the isovector nucleon OP given by the
isovector density dependence (\ref{c1}) of the CDM3Y6 interaction in comparison
with the JLM results \cite{Je77} at $E=35$ and 45 MeV.} \label{f4}
\end{figure*}
Using Eq.~(\ref{c2}), the isovector part of the \emph{real} nucleon OP in the
NM limit is given explicitly as
\begin{equation}
 V_{\rm IV}(E,\rho)=F^{\rm V}_{\rm IV}(E,\rho)\{(\rho_n-\rho_p)J^{\rm D}_{01}
 +\int[\rho_n\hat{j_1}(k^n_Fr)
 -\rho_p\hat{j_1}(k^p_Fr)]v^{\rm EX}_{01}(r)j_0(kr)d^3r\}.
 \label{c4}
\end{equation}
Here $J^D_{01}=\displaystyle\int v^{\rm D}_{01}(r)d^3r,\
 \hat j_1(x)=3j_1(x)/x$ and $j_1(x)$ is the first-order spherical Bessel
function; $\rho_n$ and $\rho_p$ are the neutron and proton densities of
asymmetric NM with a total density $\rho=\rho_n+\rho_p$ and the corresponding
Fermi momenta $k^{p(n)}_F=(3\pi^2\rho_{p(n)})^{1/3}$. The momentum $k$ of the
incident nucleon of mass $m$ is determined self-consistently \cite{Kho93} from
the nucleon incident energy $E$ and real OP as
\begin{equation}
k=\sqrt{\frac{2m}{\hbar^2}\{E-[V_{\rm IS}(E,\rho)\pm V_{\rm IV}(E,\rho)]\}}.
 \label{c5}
\end{equation}
Here $V_{\rm IS}(E,\rho)$ is the isoscalar part of the \emph{real} nucleon OP,
the (+) sign pertains to incident neutron and (-) sign to incident proton. Due
to the self-consistent definition (\ref{c5}) of the momentum $k$, the isovector
potential (\ref{c4}) is obtained by an iterative procedure. After $V_{\rm
IV}(E,\rho)$ is determined, the isovector part
 $W_{\rm IV}(E,\rho)$ of the \emph{imaginary} nucleon OP is obtained
from the same Eq.~(\ref{c4}), but with $F^{\rm V}_{\rm IV}(E,\rho)$ replaced by
$F^{\rm W}_{\rm IV}(E,\rho)$.
\begin{figure*}[htb] \vspace*{-0.5cm}
\mbox{\epsfig{file=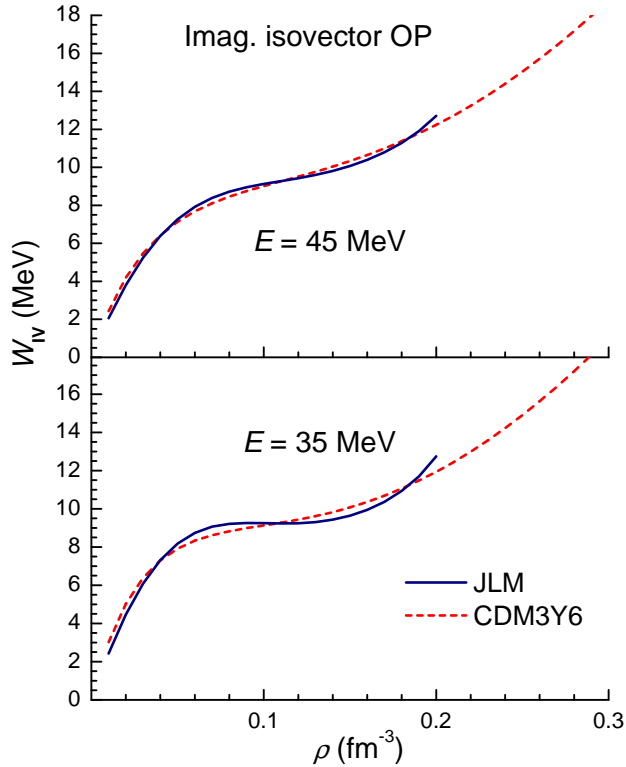,height=13cm}}\vspace*{-1.5cm}
 \caption{(Color online) the same as Fig.~\ref{f4} but for the imaginary part
$W_{\rm IV}(E,\rho)$ of the isovector nucleon OP.} \label{f5}
\end{figure*}
Our approach is to find realistic parameters of the isovector density
dependence (\ref{c1}) of the CDM3Y6 interaction by a $\chi^2$-fitting procedure
which gives the isovector part of the nucleon OP as close as possible to that
of the JLM nucleon optical potential $V^{\rm JLM}_{\rm IV}(E,\rho)$ tabulated
in Ref.~\cite{Je77}. To keep a good accuracy of this fitting procedure, instead
of introducing an energy dependent scaling factor like $g(E)$ in
Eq.~(\ref{g1}), the density dependent parameters in Eq.~(\ref{c1}) have been
adjusted separately at each energy. As illustration, the real $V_{\rm
IV}(E,\rho)$ and imaginary $W_{\rm IV}(E,\rho)$ parts of the isovector nucleon
OP at 35 and 45 MeV given by the best-fit density dependent parameters
(\ref{c1}) are compared with the JLM results \cite{Je77} in Figs.~\ref{f4} and
\ref{f5}, respectively. For each target nucleus, the parameters of complex
isovector density dependence have been searched individually at the effective
incident energy $E=E_p-Q/2$ for the calculation of the \pn form factor
(\ref{e3b}) and OP in the outgoing channel (\ref{e6b}). In all cases, the
isovector nucleon OP given by the best-fit parameters agree closely with the
JLM results in the same way as shown in Figs.~~\ref{f4} and \ref{f5} for $E=35$
and 45 MeV. The numerical parameters of isovector density dependence (\ref{c1})
at different energies $E$ can be obtained from the authors upon request. For
the HF calculation of nuclear matter, the isovector density dependence
(\ref{c1}) of the CDM3Y6 interaction at energy $E$ approaching zero has also
been constructed based on the JLM results \cite{Lej80} for low energies ($0< E
<10$ MeV). This set of density dependent parameters is used in the present work
to calculate the density dependent NM symmetry energy $S(\rho)$, defined in
Eq.~(\ref{e2}), by the HF method developed in Ref.~\cite{Kho96} explicitly for
use with the isospin- and density dependent M3Y interaction.

\begin{figure*}[htb] \vspace*{-2cm}
\mbox{\epsfig{file=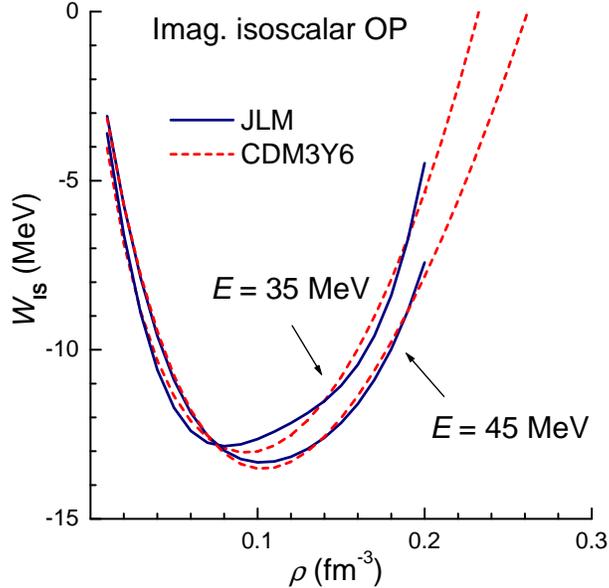,height=13.5cm}}\vspace*{-3cm} \caption{(Color
online) imaginary part $W_{\rm IS}(E,\rho)$ of the isoscalar nucleon OP given
by the isoscalar density dependent interaction (\ref{c6}) in comparison with
the JLM results \cite{Je77} at $E=35$ and 45 MeV.} \label{f5a}
\end{figure*}

In the context of a fully microscopic OP, it is also desirable to have a
realistic \emph{imaginary} isoscalar density dependence for use in the folding
calculation with the real isoscalar density dependence (\ref{g1}) of the CDM3Y6
interaction. Thus, we define the imaginary isoscalar interaction based on the
same density dependent functional (\ref{c1}) as
\begin{equation}
 w^{\rm D(EX)}_{00}(E,\rho,s)=F^{\rm W}_{\rm IS}(E,\rho)v^{\rm D(EX)}_{00}(s),
 \label{c6}
\end{equation}
then the imaginary isoscalar nucleon OP in the nuclear limit is given by
\begin{equation}
 W_{\rm IS}(E,\rho)=F^{\rm W}_{\rm IS}(E,\rho)\{\rho J^{\rm D}_{00}
 +\int[\rho_n\hat{j_1}(k^n_Fr)+\rho_p\hat{j_1}(k^p_Fr)]
 v^{\rm EX}_{01}(r)j_0(kr)d^3r\}.
 \label{c7}
\end{equation}
Here $J^D_{00}=\displaystyle\int v^{\rm D}_{00}(r)d^3r$, and other involved
variables are determined in the same way as those in Eq.~(\ref{c4}). In a
similar manner, the parameters of $F^{\rm W}_{\rm IS}(E,\rho)$ have been
searched at each energy to reproduce the JLM results tabulated in
Ref.~\cite{Je77}. As an example, the isoscalar potential $W_{\rm IS}(E,\rho)$
given by the best-fit parameters and the corresponding JLM potential at $E=35$
and 45 MeV are shown in Fig.~\ref{f5a}.
We must note that the imaginary OP based on the JLM results for nuclear matter
describes the absorption due to to the Pauli blocking effect which leads to a
\emph{finite} mean-free path of nucleons in nuclear medium. As a result,
$W_{\rm IS}(E,\rho)$ tends to have a volume shape (deep in the center and
shallow at the surface). In general, the imaginary nucleon OP at low and medium
energies has been found \cite{Va91,Kon03} to be best represented by a
combination of volume and surface terms. The surface absorption is caused
mainly by the collective excitations and transfer reactions which occur in the
nuclear surface and are not related to the ``volume" absorption given by
$W_{\rm IS}(E,\rho)$.

In conclusion, we have used the HF method (\ref{c4})-(\ref{c7}) to construct a
complex isovector density dependence of the CDM3Y6 interaction based on the JLM
results for the nucleon OP in nuclear matter \cite{Je77}. In connection with
our study, we recall that the original version of the JLM interaction was
simply deduced from the JLM nucleon OP in nuclear matter using a local density
approximation and smeared out in the coordinate space by a single Gaussian
\cite{Je77,Bau98}. For example, the real part of the original JLM interaction
was constructed by this method from the real nucleon OP in nuclear matter
$V^{\rm JLM}_{\rm IS(IV)}(E,\rho)$ as
\begin{equation}
 v_{00(01)}(E,\rho,s)\sim
 \frac{V^{\rm JLM}_{\rm IS(IV)}(E,\rho)}{\rho}\exp(-\frac{s^2}{t^2}),
 \label{c8}
\end{equation}
with the Gaussian range $t$ chosen to give a good global fit to the elastic
data. Since $V^{\rm JLM}_{\rm IS(IV)}(E,\rho)$ already contains strengths of
both direct and exchange parts of the G-matrix, the \nA OP for finite nuclei is
given by the direct folding integration (\ref{e10}) only. Despite the
simplicity, the original JLM interaction has been used quite successfully to
study the elastic \nA scattering \cite{Bau98} as well as the \pn reaction to
IAS \cite{Pak01,Bau01}.

\subsection{Results and discussions}
Given the new complex density dependence of CDM3Y6 interaction, the isoscalar
and isovector parts of the \nA OP can be calculated explicitly by the
single-folding approach (\ref{e10})-(\ref{e11}). It is natural, as the first
step, to check the OM description of elastic \nA scattering at the nearby
energies using the complex microscopic OP
\begin{equation}
 U(R)=N_{\rm V}[V_{\rm IS}(R)\pm V_{\rm IV}(R)]+
 iN_{\rm W}[W_{\rm IS}(R)\pm W_{\rm IV}(R)],
\label{g6a}
\end{equation}
where the (+) sign pertains to incident neutron and (-) sign to incident
proton. Note that the imaginary part $W_{\rm IS(IV)}(R)$ of the OP is given by
the same folding procedure (\ref{e10})-(\ref{e11}) but using the imaginary
parts (\ref{c3}) and (\ref{c6}) of the CDM3Y6 interaction constructed
separately at each energy. $U$ is further added by the spin-orbital potential
(and \pA OP added also by the Coulomb potential) taken, for simplicity, from
the CH89 systematics \cite{Va91}. The strengths $N_{\rm V(W)}$ of the complex
folded OP are adjusted to the best OM fit to the elastic scattering data.
\begin{figure*}[htb]
 \vspace*{-1cm}
 \mbox{\epsfig{file=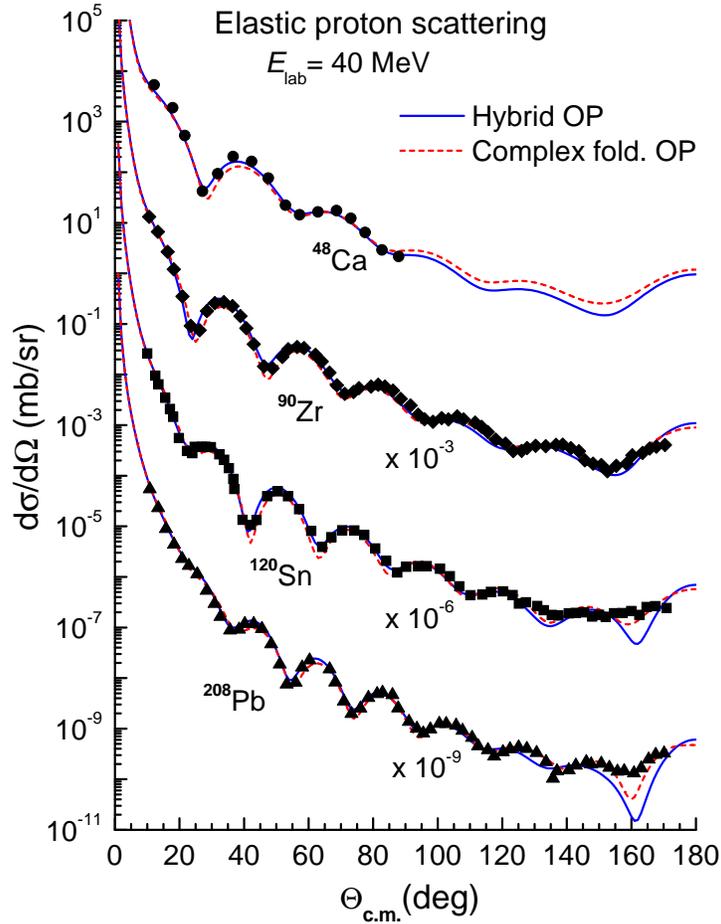,height=15cm}}\vspace*{-1cm}
\caption{(Color online) OM description of the elastic proton scattering from
$^{48}$Ca, $^{90}$Zr, $^{120}$Sn, and $^{208}$Pb targets at 40 MeV obtained
with the complex folded OP (\ref{g6a}) and hybrid OP (\ref{g6}). The data were
taken from Refs.~\cite{Gru72,Fri67}.} \label{f5b}
\end{figure*}
The OM results obtained for the elastic proton scattering at 40 MeV on the
targets under study are shown in Fig.~\ref{f5b}. A good description of the
measured elastic proton scattering data \cite{Gru72,Fri67} can be reached after
the complex folded potential is renormalized by $N_{\rm V}\approx 0.90-0.94$
and $N_{\rm W}\approx 0.6-0.8$. The OM results obtained for the elastic neutron
scattering are shown in Fig.~\ref{f5c} where the best-fit $N_{\rm V}\approx
0.9$ and $N_{\rm W}\approx 0.6-0.7$.
\begin{figure*}[htb]
 \vspace*{-1cm}
 \mbox{\epsfig{file=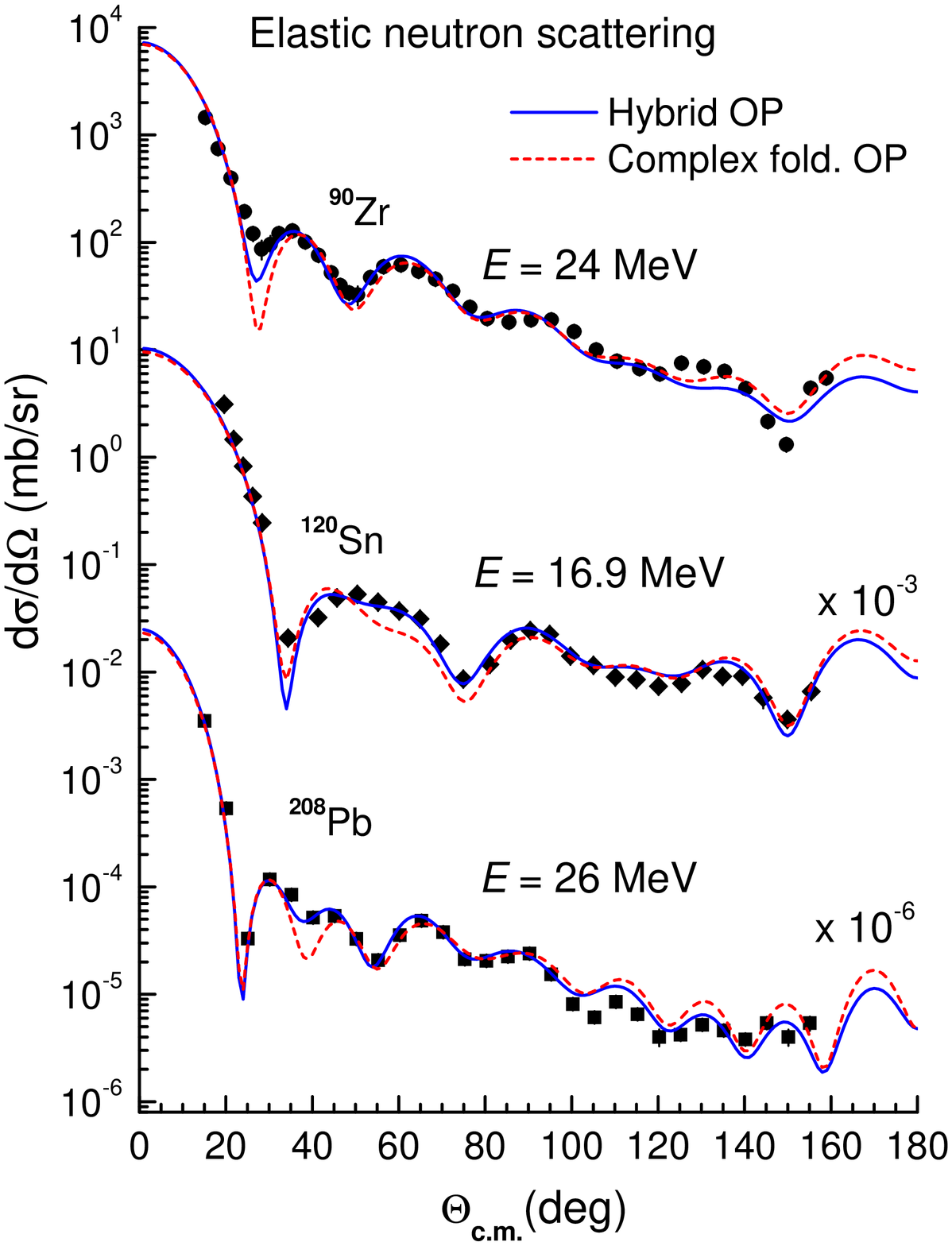,height=15cm}}\vspace*{-1cm}
\caption{(Color online) OM description of the elastic neutron scattering from
$^{90}$Zr, $^{120}$Sn, and $^{208}$Pb targets at energies of 17 to 24 MeV
obtained with the complex folded OP and (\ref{g6a}) and hybrid OP (\ref{g6}).
The data were taken from Refs.~\cite{Wan90,Gus89,Rap78}.} \label{f5c}
\end{figure*}
To have accurate distorted waves for the CC study of \pnA reaction, we have
also tried a hybrid choice for the complex OP with its real part given by the
folding model and imaginary part by a Woods-Saxon (WS) potential based on the
CH89 global systematics \cite{Va91}
\begin{eqnarray}
 U(R) & = & N_V[V_{\rm IS}(R)\pm V_{\rm IV}(R)]
 -i[W_vf(R)-4a_wW_s\frac{df(R)}{dR}], \nonumber \\
\ {\rm where}\ f(R) & = & 1/\{1+\exp[(R-R_w)/a_w\}. \label{g6}
\end{eqnarray}
The normalization factor $N_{\rm V}$ of the real folded potential as well as
strengths of the volume ($W_{\rm v}$) and surface ($W_{\rm s}$) terms of the
absorptive WS potential are fine tuned in each case to fit the elastic
scattering data under study and/or reproduce in the OM calculation the total
reaction cross section $\sigma_{\rm R}$ measured for the considered \pA systems
at 35 and 45 MeV \cite{Ca96}. The OM descriptions of the elastic proton and
neutron scattering given by such a hybrid choice for the nucleon OP are shown
in Figs.~\ref{f5b} and \ref{f5c} as solid curves. One can see that the OM fit
given by the hybrid OP is slightly improved, especially, at forward scattering
angles. Although the difference in the OM description of elastic nucleon
scattering by the two choices of OP is marginal as seen in Figs.~\ref{f5b} and
\ref{f5c}, their effect on the calculated \pn cross section is surprisingly
much more significant.

After the OP for the entrance \pA channel is determined based on the OM
analysis of the proton elastic scattering at the nearby energies, the (complex)
charge exchange FF for the \pn transition channel is determined from the real
and imaginary parts of the folded isovector potential (\ref{e11}), evaluated at
$E=E_{\rm lab}-Q/2$, as
\begin{eqnarray}
 F_{pn}(R)=\frac{2}{A}\sqrt{2T_z}U_1(R)=\sqrt{\frac{2}{T_z}}
 [N_{\rm R}V_{\rm IV}(R)+iN_{\rm I}W_{\rm IV}(R)]. \label{g7}
\end{eqnarray}
Keeping the OP parameters unchanged as fixed from the OM calculation described
above, the normalization factors $N_{\rm R(I)}$ of the folded charge exchange
FF were adjusted for the best fit of the calculated \pn cross section to the
data. In this way, the folding model analysis of the \pn reaction can serve as
a good probe of the isospin dependence of the effective $NN$ interaction. Since
the elastic neutron scattering on a target being in its \emph{excited} IAS
cannot be measured, the complex OP for the outgoing \An channel has been
determined from the complex proton OP evaluated at the effective incident
energy $E=E_{\rm lab}-Q/2$, based on the isospin coupling (\ref{e6b}). For
consistency, the complex folded OP in the \An channel is renormalized by the
same factors $N_{\rm V(W)}$ as those used in entrance \pA channel. The WS
imaginary part of the hybrid OP (\ref{g6}) in the outgoing \An channel is
determined from the CH89 global OP using the same isospin coupling (\ref{e6b}).
The OP parameters used in our CC calculation of the \pn reaction are given in
Tables~\ref{t2} and \ref{t3} for the complex folded and hybrid OP,
respectively.

We discuss now in details the CC results for the \pn reaction measured with
$^{48}$Ca target.
\begin{figure*}[htb] \vspace*{-0.5cm}
\mbox{\epsfig{file=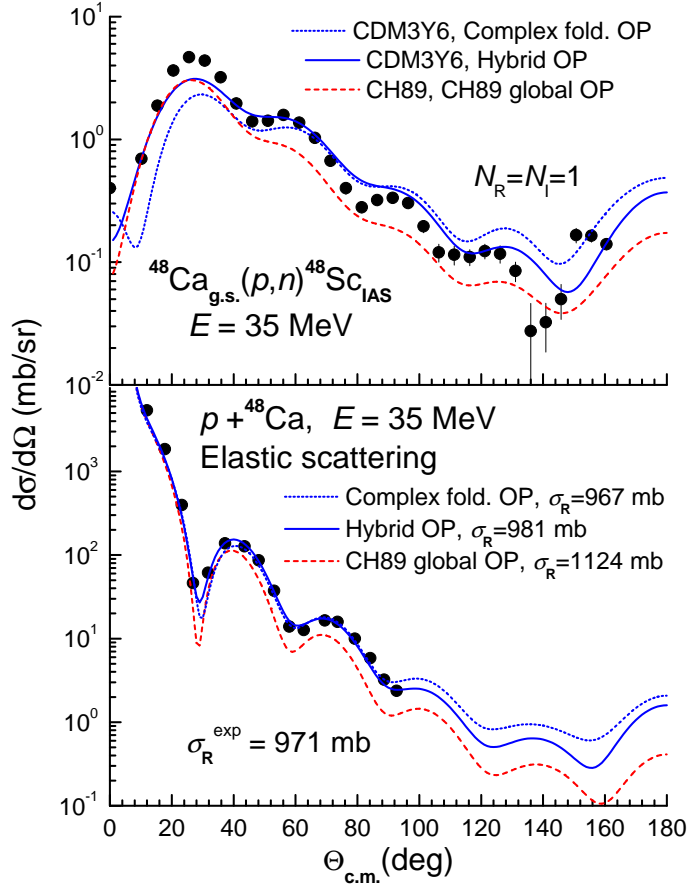,height=14cm}}\vspace*{-1cm} \caption{(Color online)
upper part: CC description of the \pnCa reaction at 35 MeV \cite{Doe75} given
by the (unrenormalized) folded \pn form factor (\ref{g7}) and that deduced from
Eq.~(\ref{e3b}) using the CH89 parameters \cite{Va91}. Lower part: OM
description of the elastic \pCa scattering at 35 MeV \cite{Gru72} given by the
complex folded OP (\ref{g6a}), hybrid OP (\ref{g6}) and CH89 global OP
\cite{Va91}.} \label{f6}
\end{figure*}
The OM descriptions of the elastic \pCa scattering data at 35 MeV \cite{Gru72}
given by the complex folded OP (\ref{g6a}), hybrid OP (\ref{g6}) and CH89
global OP \cite{Va91} are shown in lower part of Fig.~\ref{f6}. Similar to the
results at 40 MeV shown in Fig.~\ref{f5b}, both complex folded and hybrid OP
give a reasonable description of the measurement after their strengths were
adjusted by the OM fit to the elastic data, with the complex folded OP slightly
underestimating data at the forward angles. The CH89 global OP \cite{Va91}
sizably underestimates the elastic scattering data, and this is caused by a
stronger absorption given by the CH89 imaginary OP. The CC descriptions of the
\pnCa data at 35 MeV \cite{Doe75} given by the unrenormalized folded form
factor (\ref{g7}) and that deduced from the isovector term of the CH89
potential using Eq.~(\ref{e3b}) are shown in upper part of Fig.~\ref{f6}, where
the same OP's as those used in the lower part of Fig.~\ref{f6} were taken for
the entrance channel and the corresponding $U_n$ potentials evaluated at
$E=E_{\rm lab}-Q/2$ taken for the outgoing channel. One can see that the
unrenormalized folded FF gives a reasonable description of the measured \pn
cross section at large angles while underestimates the data points at the
forward angles. From the two choices of the OP, the complex folded OP
(\ref{g6a}) gives a worse fit to the \pn data at forward angles. Since the
angular distribution at forward angles is more affected by the surface part of
the OP and given the same real folded OP used in both calculations, the
difference caused by the two OP's should be due to different surface
absorptions described by the two OP's. The role of absorption is also seen in
the CC description of the \pn data by the Lane FF determined from the CH89
parameters (denoted hereafter as CH89 form factor). Namely, the CH89 form
factor sizably underestimates the data over the whole angular range when the
OP's in the entrance and outgoing channels are taken exactly as given by the
CH89 systematics \cite{Va91}. The CC description by the CH89 form factor
improves significantly when the best-fit hybrid OP (\ref{g6}) is used (see
Fig.~\ref{f7}). Therefore, the unsatisfactory description of the \pn data by
the CH89 form factor shown in upper part of Fig.~\ref{f6} is caused by a too
absorptive imaginary CH89 potential (which gives $\sigma_{\rm R}\approx 1124$
mb compared to the measurement $\sigma^{\rm exp}_{\rm R}\approx 971\pm 32$ mb
\cite{Ca96}).
\begin{figure*}[htb] \vspace*{-0.5cm}
\mbox{\epsfig{file=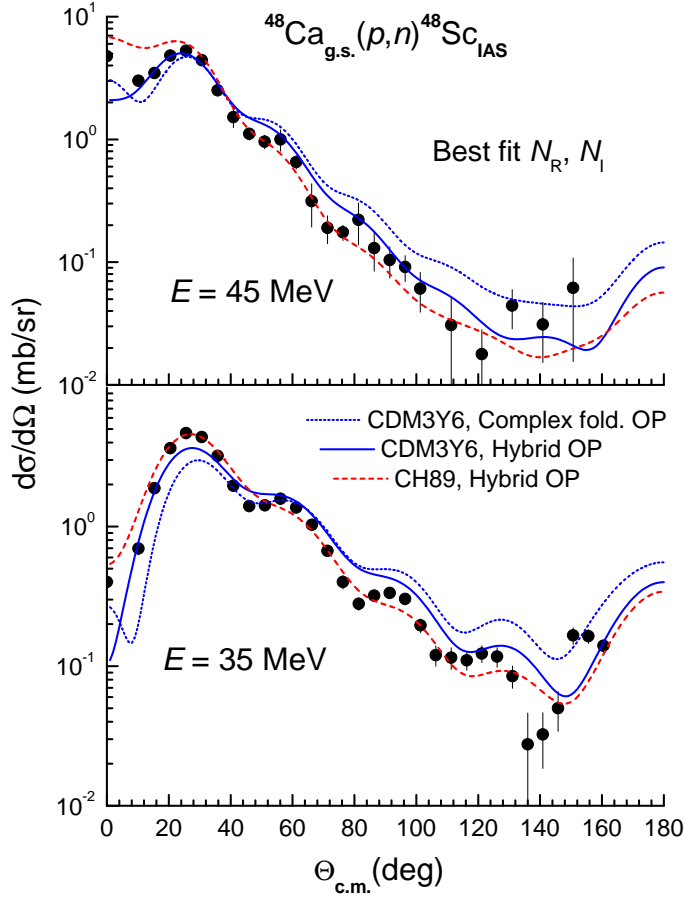,height=14cm}}\vspace*{-1cm} \caption{(Color
online) CC description of the \pnCa reaction \cite{Doe75} at 35 MeV (lower
part) and 45 MeV (upper part) given by the renormalized folded \pn form factor
(\ref{g7}) and that deduced from Eq.~(\ref{e3b}) using CH89 global OP
\cite{Va91}, using two choices (\ref{g6a})-(\ref{g6}) of the OP.} \label{f7}
\end{figure*}
We have further adjusted the complex strength of the folded FF to the best
$\chi^2$-fit of the \pn data at 35 MeV \cite{Doe75}, and $N_{\rm R}$ turns out
to be around 1.3, while $N_{\rm I}$ remains close to unity (see lower part of
Fig.~\ref{f7} and Tables~\ref{t2} and \ref{t3}). The deficiency of the complex
folded OP cannot be eliminated by such an adjustment of the folded FF. A
consistency check has also been made with all the folding and CC calculations
redone using the real isovector component of CDM3Y6 interaction increased by a
factor of 1.3, and all the calculated cross sections remain about the same (as
seen in the logarithmic graphing). The effect by the imaginary OP becomes more
substantial in the CC analysis of the \pn data at 45 MeV (upper part of
Fig.~\ref{f7}). While the use of the hybrid OP (\ref{g6}) results on about the
same best-fit $N_{\rm R(I)}$ coefficients of the folded FF as those found at 35
MeV, the complex folded OP (\ref{g6a}) gives a much larger $N_{\rm R}$ of
around 1.7 and a worse description of the \pn data at large angles.
\begin{figure*}[htb]\vspace*{-0.5cm}
\mbox{\epsfig{file=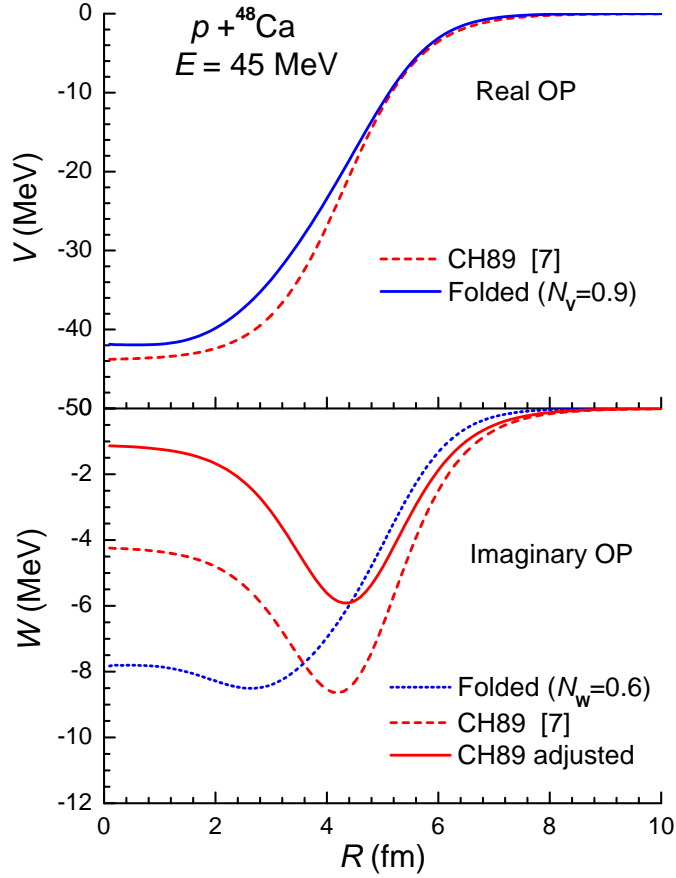,height=14cm}}\vspace*{-1cm} \caption{(Color
online) the real and imaginary parts of the complex folded OP (\ref{g6a}) for
\pCa system at 45 MeV, scaled by factors $N_{\rm V(W)}$ given by the OM fit of
the elastic data at 40 MeV, in comparison with the CH89 global OP taken from
Ref.~\cite{Va91}. The adjusted imaginary CH89 potential has been used in the
hybrid choice (\ref{g6}) of the OP.} \label{f7a}
\end{figure*}
Since the real OP and complex folded FF are exactly the same in both
calculations, the difference in the calculated \pn cross sections (solid and
dotted lines in Fig.~\ref{f7}) is entirely due to the difference in the
imaginary OP's. For illustration, we have plotted radial shapes of the \pCa
optical potential at 45 MeV in Fig.~\ref{f7a}. One can see that the real folded
potential (renormalized by $N_{\rm V}\approx 0.9$) is quite close in strength
and shape to the global CH89 real OP. The situation with the imaginary OP is
quite different: while the imaginary folded potential has a volume shape, the
imaginary CH89 potential is strongly surface peaked even after its strengths
$W_{\rm v}$ and $W_{\rm s}$ were reduced by the OM fit to the elastic proton
scattering data. The obvious reason is that the imaginary folded potential
(\ref{g6a}) has been constructed based on the imaginary nucleon OP given by the
BHF calculation of NM and is, therefore, of a ``volume" nature. As a result,
the imaginary folded potential cannot properly account for the surface
absorption caused by inelastic scattering to the low-lying collective
excitations and transfer reactions. The renormalization factor of the folded FF
was driven to the excessive value of $N_{\rm R}\approx 1.7$ by the use of the
imaginary folded potential and \emph{not} by the weakness of the isovector
interaction (\ref{c2}) - (\ref{c3}). We note further that this subtle
``absorption" effect has been established only in the CC calculation of the \pn
reaction to IAS because the elastic nucleon scattering data at the nearby
energies are still reasonably reproduced with the imaginary folded potential
(see Figs.~\ref{f5b} and \ref{f5c}). Thus, the distorted waves $\chi_{pA}$ and
$\chi_{n\tilde{A}}$ given by the realistic hybrid OP (\ref{g6}) should be more
accurate for the probe of the isovector density dependence in the CC analysis
of the \pn reaction. The CC calculations using the hybrid OP (\ref{g6}) give a
good overall description of the \pnCa data at 35 and 45 MeV with the folded FF
renormalized by $N_{\rm R}\approx 1.3$ and $N_{\rm I}\approx 1$. These
calculations also give the total \pn cross section $\sigma_{pn}\approx 10.7$
and 9.0 mb for the \pnCa reaction at 35 and 45 MeV, respectively, which agree
well with the measured values \cite{Doe75}, $\sigma_{pn}^{\rm exp}\approx
10.2\pm 1.1$ and $8.4\pm 1.0$ mb at 35 and 45 MeV, respectively.
\begin{table}
\caption{Renormalization coefficients $N_{\rm V(W)}$ of the complex folded \pA
OP (\ref{g6a}) used in the entrance channel. The calculated proton total
reaction cross section $\sigma_{\rm R}$ is compared with the data
 $\sigma_{\rm R}^{\rm exp}$ taken from Ref.~\cite{Ca96}. $N_{\rm R(I)}$ are
the renormalization coefficients of the folded FF (\ref{g7}) deduced from the
CC fit to the \pn data using the OP (\ref{g6a}).} \label{t2}
\begin{tabular}{|c|c|c|c|c|c|c|c|} \hline
 Target & $E$ & $N_{\rm V}$ & $N_{\rm W}$ & $\sigma_{\rm R}$ &
 $\sigma_R^{\rm exp}$ & $N_{\rm R}$ & $N_{\rm I}$ \\
 $A$ & (MeV) & & & (mb) & (mb) &  & \\ \hline
 $^{48}$Ca & 35 & 0.933 & 0.600 & 969 & $971\pm 32$ & 1.356 & 0.970 \\
  & 45 & 0.902 & 0.630 & 893 & $908\pm 34$ & 1.738 & 1.054 \\
$^{90}$Zr & 35 & 0.893 & 0.731 & 1341 & $1316\pm 65\ ^a$ & 2.133 & 0.978 \\
  & 45 & 0.893 & 0.731 & 1296 & $1214\pm 59\ ^b$ & 2.193 & 1.043 \\
  $^{120}$Sn & 35 & 0.937 & 0.828 & 1605 & $1668\pm 59$ & 2.372 & 0.981 \\
  & 45 & 0.937 & 0.731 & 1588 & $1545\pm 38$ & 2.529 & 0.985 \\
  $^{208}$Pb & 35 & 0.916 & 0.747 & 1877 & $1974\pm 38$ & 2.896 & 1.018 \\
  & 45 & 0.916 & 0.747 & 1963 & $1979\pm 41$ & 2.606 & 0.985 \\
  \hline
\end{tabular}

$^a$ Total \pZr reaction cross section measured at $E=40$ MeV;
$^b$ at $E=49.5$ MeV. \\
\end{table}

\begin{table}
\caption{Parameters of the hybrid OP (\ref{g6}) used in the entrance and exit
channels. Parameters given in boldface were kept unchanged as determined from
the CH89 systematics \cite{Va91}. The calculated proton total reaction cross
section $\sigma_{\rm R}$ is compared with the data $\sigma_{\rm R}^{\rm exp}$
taken from Ref.~\cite{Ca96}. $N_{\rm R(I)}$ are the renormalization
coefficients of the folded FF (\ref{g7}) deduced from the CC fit to the \pn
data using the OP (\ref{g6}).} \label{t3}
\begin{tabular}{|c|c|c|c|c|c|c|c|c|c|c|c|} \hline
 Target & $E$ & Channel & $N_{\rm V}$ & $W_v$ & $W_s$ & $R_w$ & $a_w$ &
 $\sigma_{\rm R}$ & $\sigma_R^{\rm exp}$ & $N_{\rm R}$ & $N_{\rm I}$ \\
 $A$ & (MeV) & & & (MeV) & (MeV) & (fm) & (fm) & (mb) &
 (mb) &  & \\ \hline
 $^{48}$Ca & 35 & \Ap & 0.925 & 1.495 & 5.432 & {\bf 4.414} &
 {\bf 0.69} & 981 & $971\pm 32$ & 1.265 & 0.960 \\
  & & \An & 0.925 & 1.495 & {\bf 4.503} & {\bf 4.414} & {\bf 0.69} &
  - & - & - & - \\
  & 45 & \Ap & 0.900 & 1.096 & 5.358 & {\bf 4.414} & {\bf 0.69} &
  893 & $908\pm 34$ & 1.279 & 0.970 \\
  & & \An & 0.900 & 1.096 & {\bf 3.985} & {\bf 4.414} & {\bf 0.69} &
  - & - & - & - \\
 $^{90}$Zr & 35 & \Ap & 0.913 & 1.479 & 6.060 & {\bf 5.540} &
  {\bf 0.69} & 1330 & $1316\pm 65\ ^a$ & 1.202 & 0.969 \\
   &  & \An & 0.913 & {\bf 1.891} & {\bf 5.267} & {\bf 5.540} &
   {\bf 0.69} & - & - & - & - \\
  & 45 & \Ap & 0.913 & 2.434 & 5.314 & {\bf 5.540} & {\bf 0.69} &
   1296 & $1214\pm 59\ ^b$ & 1.298 & 1.081 \\
  & & \An & 0.913 & {\bf 2.918} & {\bf 4.721} & {\bf 5.540} & {\bf 0.69} &
  - & - & - & - \\
$^{120}$Sn & 35 & \Ap & 0.937 & {\bf 2.305} & {\bf 7.792} & {\bf 6.140}
 & {\bf 0.69} & 1637 & $1668\pm 59$ & 1.203 & 0.950 \\
   & & \An & 0.937 & {\bf 1.686} & {\bf 4.687} & {\bf 6.140} & {\bf 0.69}
    & - & - & - & - \\
   & 45 & \Ap & 0.937 & 2.027 &  6.529 & {\bf 6.140} & {\bf 0.69}&
  1570 & $1545\pm 38$ & 1.225 & 0.958 \\
 &  & \An & 0.937 & {\bf 2.653} & {\bf 4.218} & {\bf 6.140} & {\bf 0.69}
 & - & - & - & - \\
$^{208}$Pb & 35 & \Ap & 0.901 & 2.419 & {\bf 8.729} & {\bf 7.460}
 & {\bf 0.69} & 1964 & $1974\pm 38$ & 1.201 & 0.955 \\
   & & \An & 0.901 & {\bf 1.127} & {\bf 4.386} & {\bf 7.460} & {\bf 0.69}
    & - & - & - & - \\
   & 45 & \Ap & 0.901 & {\bf 2.827} & 6.334 & {\bf 7.460} & {\bf 0.69}&
  1998 & $1979\pm 41$ & 1.150 & 0.930 \\
 &  & \An & 0.901 & {\bf 1.871} & {\bf 4.000} & {\bf 7.460} & {\bf 0.69}
 & - & - & - & - \\ \hline
\end{tabular}

$^a$ Total \pZr reaction cross section measured at $E=40$ MeV;
$^b$ at $E=49.5$ MeV. \\
\end{table}

\begin{figure*}[htb] \vspace*{-0.5cm}
\mbox{\epsfig{file=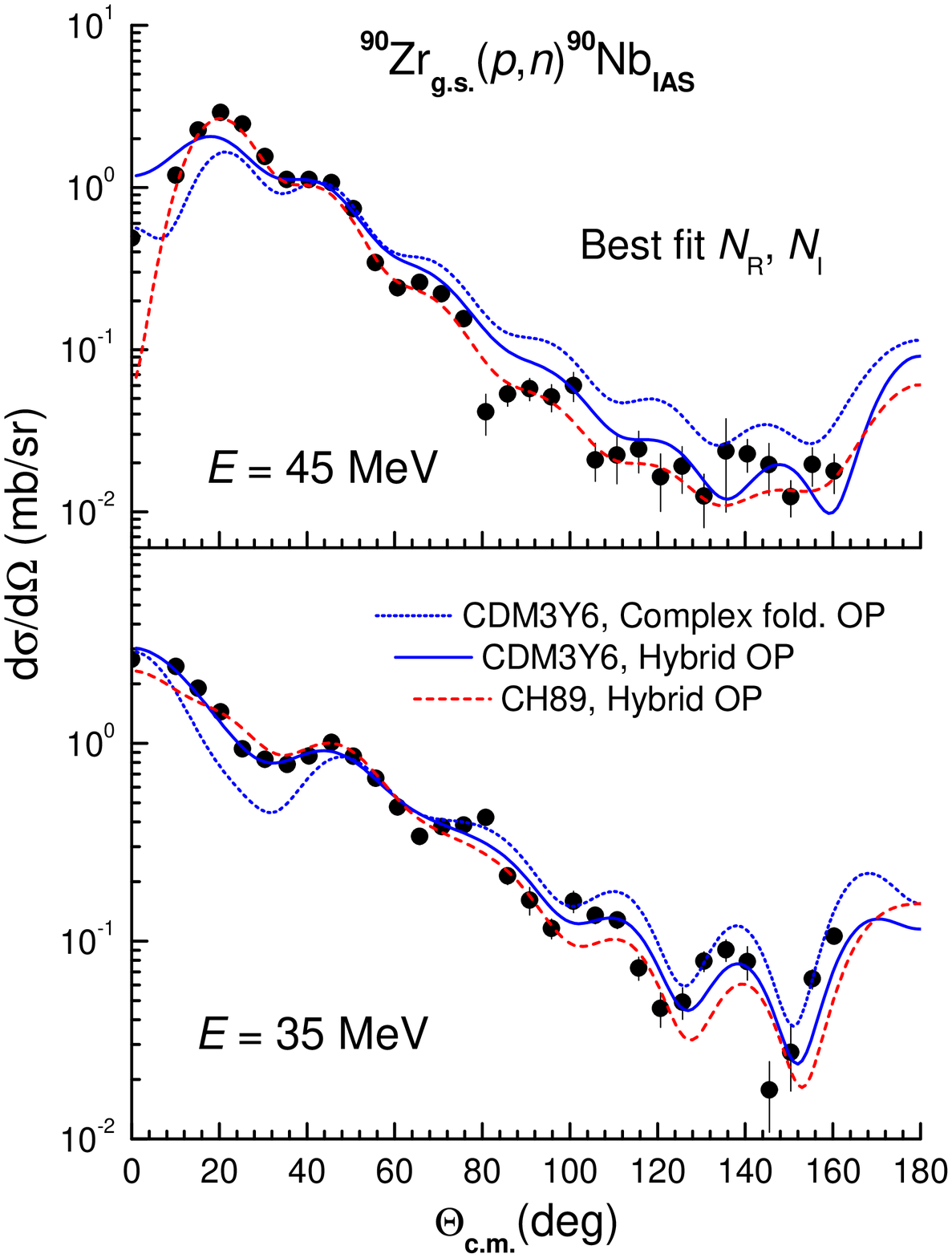,height=14cm}}\vspace*{-1cm} \caption{(Color online)
the same as Fig.~\ref{f7} but for the \pnZr reaction \cite{Doe75}.} \label{f9}
\end{figure*}

The results of our folding model analysis of the \pnZr reaction at the same
energies are compared with the data in Fig.~\ref{f9}. One can see that the peak
of the \pn cross section is weaker and only around half of that measured for
\pnCa reaction. A weaker charge exchange strength also results on the total \pn
cross section of about 50\% smaller than that obtained for $^{48}$Ca target
(see Table~I in Ref.~\cite{Doe75}). In terms of the isospin-flip transition
(\ref{e3b}), the charge exchange \pn strength is directly proportional to the
neutron-proton asymmetry parameter $\delta=(N-Z)/A$ and strength of the Lane
potential $U_1$. Indeed, the isovector folded potential $V_{\rm IV}(R)$ for the
\pCa system is about 30-40\% larger than that obtained for the \pZr system at
the surface distances and the asymmetry parameter $\delta\approx 0.17$ and 0.11
for $^{48}$Ca and $^{90}$Zr, respectively. A weaker charge exchange strength
observed in the \pnZr reaction is, therefore, well anticipated. Like the \pCa
system, the use of the complex folded OP (\ref{g6a}) in the CC calculation with
the folded FF gives a poor description of the \pn data, especially at forward
angles  (see Fig.~\ref{f9}), even after its real strength is renormalized by
$N_{\rm R}> 2$ as determined from the $\chi^2$ fit to the data. A strongly
``volume" imaginary folded potential is also the reason for this disagreement.
The same folded FF gives a much better fit to the \pn data when the hybrid OP
(\ref{g6}) is used and its complex strengths need to be renormalized by just
$N_{\rm R}\approx 1.2-1.3$ and $N_{\rm I} \approx 1$ which are close to those
obtained for the \pCa system (see Table~\ref{t3}). The CH89 form factor for the
\pZr system slightly underestimates the data if the OP in the entrance and exit
channels are determined as given by the CH89 parameters. However, the CC
description of the \pn data by the CH89 form factor is much better when the
hybrid OP (\ref{g6}) is used. The CC calculation using the hybrid OP and
renormalized folded FF gives the total \pn cross section $\sigma_{pn}=4.8$ and
4.1 mb for the \pnZr reaction at 35 and 45 MeV, respectively, which agree
nicely with the data ($\sigma_{pn}^{\rm exp}\approx 4.8\pm 0.5$ and $4.4\pm
0.5$ mb at 35 MeV and 45 MeV, respectively) \cite{Doe75}.

\begin{figure*}[htb] \vspace*{-0.5cm}
\mbox{\epsfig{file=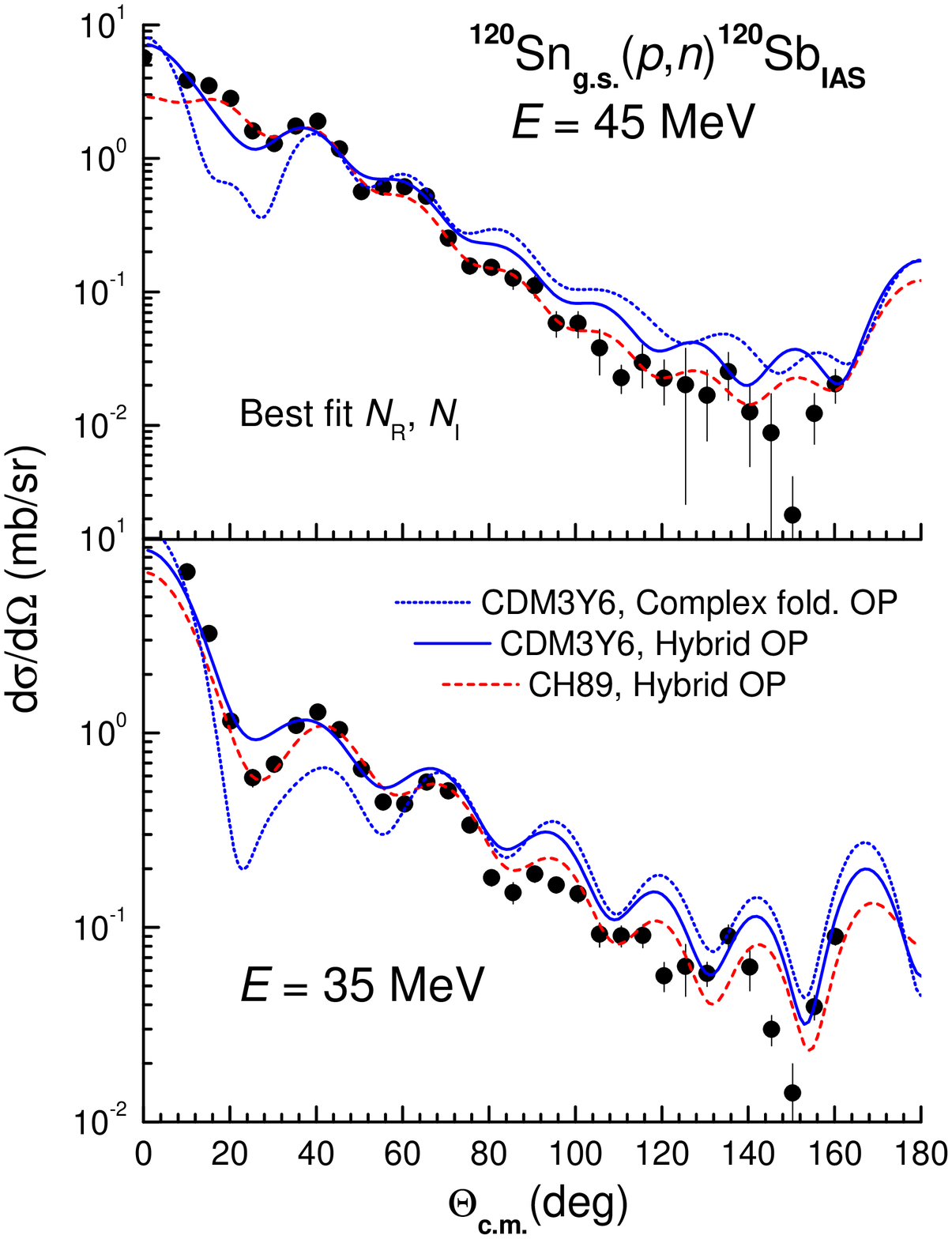,height=14cm}}\vspace*{-1cm} \caption{(Color
online) the same as Fig.~\ref{f7} but for the \pnSn reaction \cite{Doe75}.}
 \label{f10}
\end{figure*}
\begin{figure*}[htb] \vspace*{-0.5cm}
\mbox{\epsfig{file=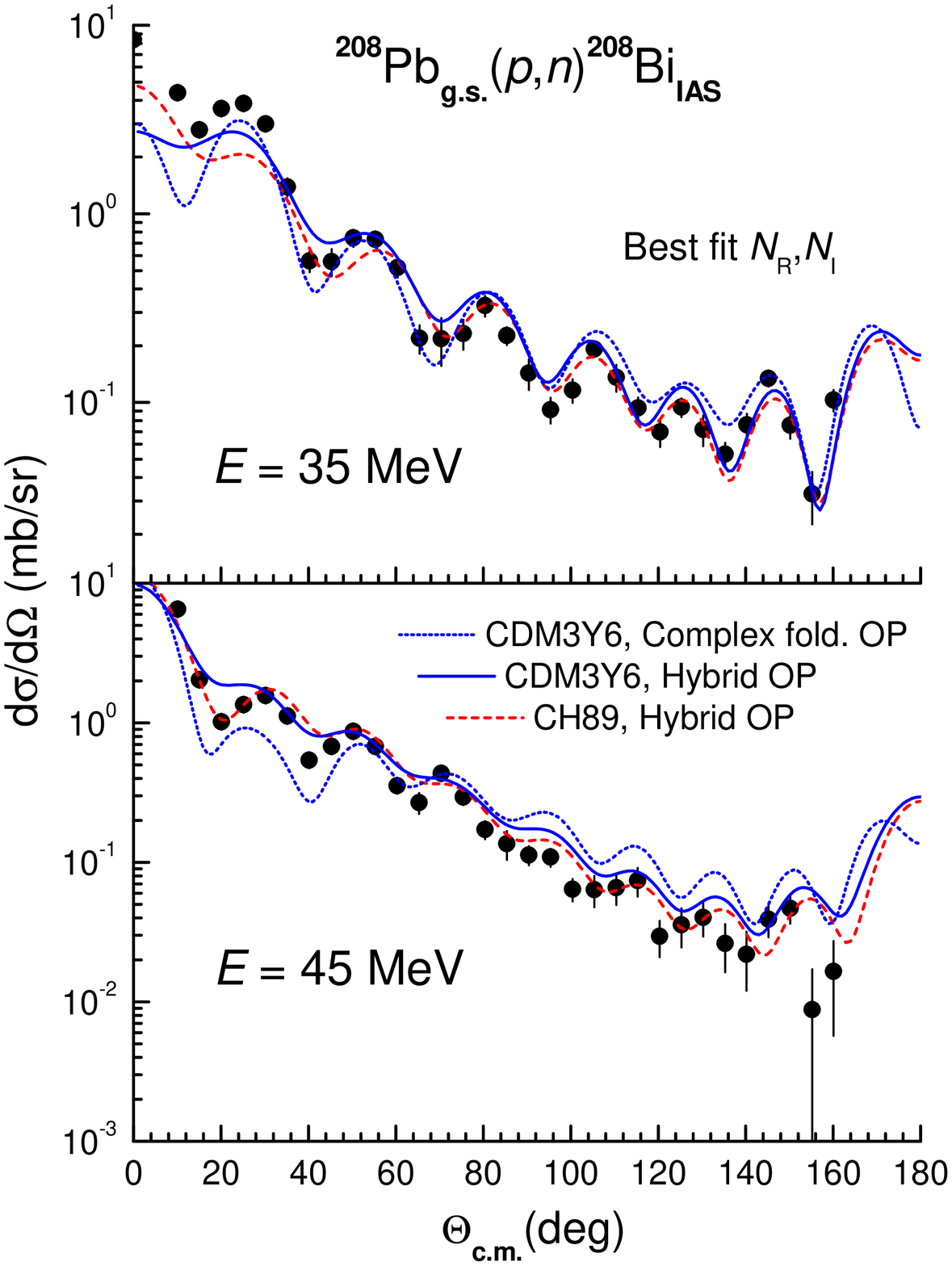,height=14cm}}\vspace*{-1cm} \caption{(Color
online) the same as Fig.~\ref{f7} but for the \pnPb reaction \cite{Doe75}.}
 \label{f11}
\end{figure*}

The CC results for the \pnSn and \pnPb reactions are presented in
Figs.~\ref{f10} and \ref{f11}, respectively. Like the results obtained above
for $^{48}$Ca and $^{90}$Zr targets, the use of the complex folded OP
(\ref{g6a}) with a volume-shape imaginary part leads to a wrong shape of the
calculated \pn cross section at forward angles (see, in particular,
Fig.~\ref{f10}). The CC description of the \pn data by both the folded FF and
CH89 form factors is very satisfactory when the hybrid OP's (which describe
well the proton elastic scattering at 40 MeV and measured total reaction cross
section) are used for the entrance and exit channels. A stronger \pA Coulomb
potential seems to push the main peak of the \pn cross section to the forward
angles (compare, e.g., Figs.~\ref{f7} and \ref{f11}), but the measured data
points in the observable angular range still allow us to make an accurate
conclusion on the complex strength of the folded \pn form factor (\ref{g7}).
For the two heavy targets, the best CC fit by the folded FF is reached when its
real and imaginary strengths are scaled by $N_{\rm R}\approx 1.2$ and $N_{\rm
I} \approx 1$ which are reasonably close to those obtained for $^{48}$Ca and
$^{90}$Zr targets.
Although the complex folded OP (\ref{g6a}) can be used to reasonably describe
the elastic proton and neutron scattering on the targets under study, the
volume absorption given by its imaginary part strongly affects the distorted
waves $\chi_{pA}$ and $\chi_{n\tilde{A}}$ at the surface and, hence, leads to a
poor description of the \pn data at forward angles and a much stronger
renormalization of the folded FF. In general, a fully Lane consistent and
accurate description of both the nucleon elastic scattering and \pn reaction to
IAS should be reached with a more accurate microscopic model for the imaginary
OP, like that developed for the nucleon OP at low energies \cite{Ba05}, capable
to take into account explicitly coupling to the most collective particle-hole
excitations in the target which lead to the surface absorption. Given such a
strong impact by the absorption to the calculated \pn cross section, the
renormalization factors of the folded \pn form factor $N_{\rm R(I)}$ obtained
with the complex folded OP should not be considered as due to the weakness of
the isovector density dependence of the CDM3Y6 interaction. We must, therefore,
rely on the best-fit $N_{\rm R(I)}$ coefficients obtained with the hybrid OP
(\ref{g6}) in deducing the strength of the isovector density dependence. Our
present results emphasize that an accurate determination of the imaginary \nA
OP is very important, especially, in the DWBA or CC studies of direct reactions
measured with unstable nuclei when elastic scattering data are not always
available.

In connection with the present study, we note that the original version of the
effective JLM interaction (\ref{c8}) has been used by Pakou {\sl et al.}
\cite{Pak01} and Bauge {\sl et al.} \cite{Bau01} to study the same \pn
reactions. The JLM folding-model analysis of the proton, neutron elastic
scattering and \pn charge exchange reaction done in Ref.~\cite{Pak01} has also
shown that the isovector part of the JLM interaction is too weak and a very
strong overall renormalization of the folded FF by $N_{\rm R}=N_{\rm I}\approx
2-2.5$ is needed to account for the measured \pn cross sections. In view of our
results obtained with the complex folded OP (\ref{g6a}), it is very likely that
such large renormalization of the folded FF has been driven by the ``volume"
absorption of the JLM complex OP used in Ref.~\cite{Pak01}. In a more elaborate
treatment of the charge exchange transition within the JLM model, Bauge {\sl et
al.} \cite{Bau01} have made the isospin coupling factor in Eq.~(\ref{e3b})
density dependent, i.e., $\sqrt{2T_z/A}=\sqrt{[\rho_n(r)-\rho_p(r)]/\rho(r)}$,
and included it into the (direct) folding integral. The JLM nucleon OP obtained
in such a density-dependent isospin coupling assumption has been thoroughly
tested in the OM analysis of the proton, neutron elastic scattering and \pn
reaction over a wide range of energies and target masses, and one can deduce
from the results shown in Fig.~1 of Ref.~\cite{Bau01} the best-fit
renormalization coefficients of the \pn folded form factor $N_{\rm R}\approx
1.5-1.6$ and $N_{\rm I}\approx 1.3-1.4$, in the energy range of $30-40$ MeV,
which are closer to our results. Despite differences in the best-fit
renormalization coefficients of the folded FF obtained in the present work and
in the JLM folding-model analyses \cite{Pak01,Bau01}, all the results show
consistently that the isovector strength of the JLM interaction is much too
weak to account for the measured \pn data. Since the isovector term of the JLM
nucleon OP has been obtained as the first-order expansion of the mass operator
of symmetric nuclear matter perturbed by a neutron excess \cite{JLM77}, a
weakness of the resulting JLM nucleon OP in asymmetric NM could well be
expected. As the charge exchange reaction to IAS is quite helpful in probing
the isospin dependence of the effective $NN$ interaction, it would be of
interest to apply similar folding model analysis to test the isospin dependence
of the nucleon OP given by more advanced BHF calculations of asymmetric NM,
like that by Zuo {\sl et al.} \cite{Zuo06}.

\begin{figure}[htb]
 \vspace*{-1cm}
 \mbox{\epsfig{file=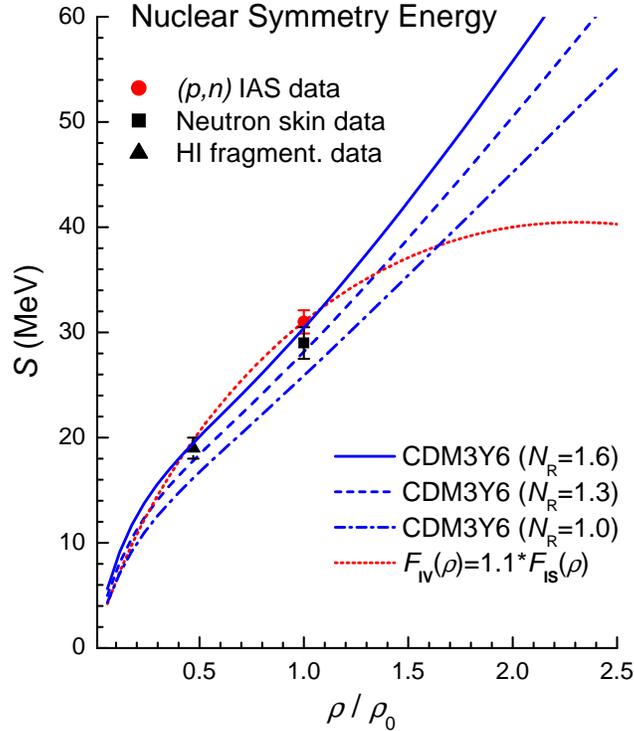,height=14cm}}\vspace*{-2cm}
\caption{(Color online) density dependence of the nuclear symmetry energy
$S(\rho)$ given by the HF calculation of asymmetric NM \cite{Kho96} using
different isovector density dependences of the CDM3Y6 interaction and the
empirical values deduced from the CC analysis of the \he6pn reaction
\cite{Kho05} as well as the neutron-skin \cite{Fur02} and HI fragmentation
\cite{Ono03,She04} studies. See more discussion in the text.} \label{f12}
\end{figure}
As mentioned above, the knowledge about the isospin dependence of the in-medium
$NN$ interaction is of vital importance in studying the equation of state of
asymmetric NM, the key input for the understanding of the dynamics of supernova
explosion and neutron star formation \cite{Bet90,Swe94,Ste05}. We show here
that the folding model analysis of the \pn reaction can be quite helpful for
the determination of the nuclear symmetry energy. After the real isovector
density dependence (\ref{c1}) of the CDM3Y6 interaction at energy approaching
zero has been carefully parameterized on the HF level to match the microscopic
BHF results \cite{Lej80}, the density- and isospin dependent CDM3Y6 interaction
is further used to calculate the nuclear symmetry energy $S(\rho)$ using the
standard HF method \cite{Kho96} and the results are shown in Fig.~\ref{f12}.
One can see that $E_{\rm sym}=S(\rho_0)$ is lying somewhat lower than the
empirical value of around $30-31$ MeV if $F^{\rm V}_{\rm IV}(\rho)$ is taken as
determined from the BHF results by JLM \cite{Je77}. The weakness of the JLM
isovector interaction is, thus, also confirmed in our HF calculation of
asymmetric NM. A very good agreement with the empirical $E_{\rm sym}$ values is
reached when $F^{\rm V}_{\rm IV}(E\approx 0,\rho)$ is scaled by a factor
$N_{\rm R}\approx 1.3-1.6$, which is slightly larger than factor $N_{\rm R}$
for $F^{\rm V}_{\rm IV}(E,\rho)$ at $E=35$ and 45 MeV deduced from our folding
model analysis of the \pn reaction. The renormalized strength of the isovector
density dependence also gives a good agreement with the empirical symmetry
energy at the half-density point, $S(\rho\approx 0.08$ fm$^{-3})\approx 18-22$
MeV, found from recent studies \cite{Ono03,She04} of the heavy-ion
fragmentation in the same energy region as that considered in our study of \pn
reactions. It should be noted that analysis of HI fragmentation data was made
based on the antisymmetrized molecular dynamics simulations \cite{Ono03} which
obtained $S(\rho\approx 0.08$ fm$^{-3})$ at a \emph{finite} temperature of
around 3 MeV. Therefore, this value approximately agrees with our HF result for
the low-density part of $S(\rho)$ shown in Fig.~\ref{f12} only if the
temperature dependence of $S(\rho)$ at low NM densities is neglected. Finally
we note that our results are also complementary to the structure studies which
relate the $E_{\rm sym}$ value to the neutron skin, a method first suggested by
Brown \cite{Bro00}. If one adopts a neutron-skin $\Delta R\approx 0.1-0.2$ fm
for $^{208}$Pb then a systematics based on the mean-field calculations
\cite{Fur02} gives $E_{\rm sym}\approx 27-31$ MeV (which is plotted as solid
square in Fig.~\ref{f12}). Although the folding model analysis of the \pn
reaction has put a constraint on the nuclear symmetry energy $S(\rho)$ at
$\rho\leq \rho_0$, its behavior at high densities remains unexplored in our
study due to a simple reason that the total nuclear density of the \pA system
never exceeds $\rho_0$, so that the \pn reaction is sensitive to the
low-density part of the isovector interaction only. In particular, when the
isovector density dependence $F_{\rm IV}(\rho)$ is taken to have the same
functional form as the isoscalar density dependence $F_{\rm IS}(\rho)$, and
scaled by a factor of 1.1 deduced from our recent CC analysis of the \he6pn
data \cite{Kho05}, it gives nearly the same description of the symmetry energy
$S(\rho)$ at $\rho\leq \rho_0$ as the newly parameterized isovector density
dependence (see dotted curve in Fig.~\ref{f12}). The two sets of the isovector
density dependence have, however, very different behaviors at high NM
densities. The $S(\rho)$ curves obtained with the isovector density dependence
based on the BHF results of asymmetric NM increase monotonically with the
increasing NM density. Such a behavior has been recently discussed as the
\emph{asy-stiff} density dependence of the nuclear symmetry energy, while the
$S(\rho)$ curve given by $F_{\rm IV}(\rho)$ having the same functional form as
$F_{\rm IS}(\rho)$ can be referred to as the \emph{asy-soft} density dependence
(see more discussion in Ref.~\cite{Bar05}). Although some HI collision data
seem to prefer the stiff density dependence of the symmetry energy
\cite{Bar05}, much more studies need to be done before a definitive conclusion
can be made. In particular, a double-folding approach to study the $(^3$He,$t$)
or $(^{13}$C,$^{13}$N) reactions exciting IAS might allow us to test the high
density part of the isovector density dependence (\ref{c1}), due to a higher
overlap nuclear density reached during the collision and, eventually, to probe
the nuclear symmetry energy $S(\rho)$ at higher NM densities.

\section{Summary}
A consistent CC analysis of the charge exchange \pn reactions to the isobaric
analog states of the ground states of $^{48}$Ca, $^{90}$Zr, $^{120}$Sn and
$^{208}$Pb targets at the proton incident energies of 35 and 45 MeV has been
done using the \pn form factors either calculated microscopically in the
folding model \cite{Kho02} or determined from the empirical WS parameters of
the existing nucleon global OP's \cite{BG69,Va91,Kon03}.

Although the isospin dependence of the CH89 global OP \cite{Va91} has been
established based only on the OM studies of the elastic proton and neutron
scattering only, it can be used to determine the charge exchange FF for the \pn
transition to IAS, based the isospin coupling (\ref{e3b}). This CH89 form
factor was shown to account quite well for the \pn data if the parameters of
the proton OP are fine tuned to reproduce the measured elastic proton
scattering and total reaction cross sections $\sigma_{\rm R}$.

To probe of the isospin dependence of the effective $NN$ interaction, a complex
isovector density dependence of the CDM3Y6 interaction \cite{Kho97} has been
carefully parameterized based on the density dependent JLM nucleon OP
\cite{Je77} and used further in the folding model analysis of the \pn reaction.
Like previous studies \cite{Pak01,Bau01} using the original JLM interaction
(\ref{c8}), the present results also show that the isovector strength of the
JLM interaction is quite weak to account for the observed \pn transitions. The
CC results obtained with realistic (semi-microscopic) nucleon OP's for the
entrance and exit channels have shown that the real isovector density
dependence needs to be enhanced by about $20-30\%$ to give a good description
of the \pn reaction.

The isovector density dependence of the CDM3Y6 interaction has also been
constructed based on the JLM nucleon OP at energy approaching zero for further
use in the HF study of asymmetric NM. The HF calculation using this new
isovector interaction gives the nuclear symmetry energy $S(\rho)$ close to the
empirical values at $\rho\leq\rho_0$ when the real isovector density dependence
is scaled by a factor $N_{\rm R}\approx 1.3-1.6$. This result confirms the
weakness of the isovector strength of the JLM interaction found in the folding
model analysis of the \pn reaction at 35 and 45 MeV. The new isovector density
dependence predicts a behavior of $S(\rho)$ at high NM densities similar to
what discussed recently in the literature \cite{Bar05} as the \emph{asy-stiff}
density dependence of the symmetry energy.

\section*{Acknowledgement}
We thank A. Pakou for her helpful communication on the \pn reactions under
study. This research project has been supported, in part, by the Natural
Science Council of Vietnam, EU Asia-Link Program CN/Asia-Link/008 (94791) and
Vietnam Atomic Energy Commission (VAEC).

\end{document}